\documentclass[10pt,prd,twocolumn,preprintnumbers,showpacs]{revtex4-1}
\pdfoutput=1
\usepackage{epsfig}
\usepackage{subfigure}
\usepackage{amssymb}
\usepackage{graphicx}
\usepackage{color}
\usepackage{amsmath}

\usepackage[
       colorlinks=true,
      filecolor=black,
       anchorcolor=blue,
      linkcolor=blue,
      citecolor=blue,
      urlcolor=blue,
       linktocpage=true,
        plainpages=false,
        breaklinks=true,
            pdfstartview=FitH
           ]{hyperref}

\def\RL{R}

\renewcommand{\thefootnote}{\fnsymbol{footnote}}

\def\be{\begin{equation}}
\def\ee{\end{equation}}
\def\ba{\begin{array}}
\def\ea{\end{array}}
\def\bea{\begin{eqnarray}}
\def\eea{\end{eqnarray}}
\def\nn{\nonumber\\}

\def\la{\label}
\def\eq#1{(\ref{#1})}

\def\a{\alpha}
\def\b{\beta}
\def\g{\gamma}

\def\d{\delta}
\def\D{\Delta}
\def\e{\epsilon}

\def\ph{\phi}

\def\l{\lambda}

\def\m{\mu}

\def\s{\sigma}

\def\o{\omega}

\def\half{\frac{1}{2}}

\def\pa{\partial}
\def\fr{\frac}

\def\bra{\left\langle}
\def\ket{\right\rangle}
\def\lb{\left[}

\def\ls{\left(}

\def\rb{\right]}

\def\rs{\right)}

\def\text#1{{\rm #1}}

\def\inf{\infty}

\begin{document}

\title{Holographic Approach to Entanglement Entropy in  Disordered Systems} 


\author{Rajesh Narayanan$^{1,2}$}\email{rnarayanan@iitm.ac.in}
\author{Chanyong Park$^{1,3,4}$}\email{cyong21@gist.ac.kr}
\author{Yun-Long Zhang$^{1,5}$}\email{yun-long.zhang@yukawa.kyoto-u.ac.jp}
 
\affiliation{}
\affiliation{$^1$Asia Pacific Center for Theoretical Physics (APCTP), Pohang  790-784, Korea} 
\affiliation{$^2$Department of Physics, Indian Institute of Technology Madras, Chennai 600036, India}
\affiliation{$^3$Department of Physics, Pohang University of Science and Technology, Pohang 790-784, Korea}
\affiliation{$^4$Department of Physics and Photon Science, Gwangju Institute of Science and Technology,  Gwangju 61005, Korea}
\affiliation{$^5$Center for Gravitational Physics, Yukawa Institute for Theoretical Physics,
Kyoto University, Kyoto  606-8502,  Japan}

\date{\small February 25, 2019}


\begin{abstract}
We investigate the entanglement entropy of a two-dimensional disordered system holographically. In particular, we study the evolution of the entanglement entropy along renormalization group flows for a conformal theory at the UV fixed point, that is perturbed by weak disorder into a Lifshitz theory at the IR fixed point.  Through numerical fitting, we find that the disorder correlations leads to a sub-leading power-law term in the entanglement entropy that vanishes at the IR fixed point. Interestingly, the exponent that controls the power-law vanishing of the sub-leading term seems to be almost universal as it depends very weakly on the strength of the disorder.  We show that our results can be put in the context of the c-theorem by defining an effective central charge that decreases along the RG flow. We also investigate disorder induced long-range correlations between the two subsystems by studying the holographic mutual information.
~~\\
~~\\
\end{abstract}



\maketitle

\renewcommand{\thefootnote}{\arabic{footnote}}
\setcounter{footnote}{0}


\allowdisplaybreaks


\section{Introduction}
\label{sec:intro}

Quenched disorder can have far reaching consequences on the physical properties of quantum materials especially near quantum critical points \cite{vojta_review}. For instance, relevant disorder fluctuations could drive the renormalization group flows away from the clean critical point \cite{harris} and lead to new critical behavior. This new critical behavior could be one which supports conventional power-law scaling with a finite value of the disorder strength \cite{harris,vojta_review} or could be of the infinite disorder kind with activated scaling \cite{fisher1,fisher2}, with accompanying exotic quantum Griffiths effects \cite{Thill_Huse:1995, Rieger_Young:1996}. In some cases disorder effects can be so stark so as to result in the ultimate destruction or smearing of the transition itself \cite{Vojta:2003,Hoyos_Vojta:2008, vojta_review, aizenmann, berker, berker2}. In addition to the effects discussed above, disorder in itself can trigger quantum phase transitions. A case in point is provided by the case of Anderson localization \cite{Anderson} wherein the wave function of quantum particles propagating in a disorder induced random potential becomes localized. 

In the recent past the Entanglement Entropy (EE) which is a good measure to characterize the Quantum Phase Transitions (QPT)\cite{SachdevQFT,osterloh,osborne,vidal, chen,amico} has been calculated for some quantum systems in the presence of quenched disorder, (see e.g. \cite{Moore2009} for a review). For instance, the Strong Disorder Renormalization (SDRG) scheme \cite{ma_dasgupta_hu,ma,fisher1,fisher2} was adapted by Refael and Moore initially to calculate the disorder averaged EE for the Random Transverse Field Ising Model and the random Heisenberg antiferromagnetic chain \cite{refael_moore}. Apart from the analytical SDRG scheme, other schemes such as the numerical exact diagonalization schemes \cite{Laflorencie2005,lin} and the Density Matrix Renormalization Group (DMRG) method \cite{chiara} have also been used to characterize the EE in disordered systems.

These results, that are centered on one dimensional disordered spin chains hosting an infinite disorder fixed point \cite{fisher1,fisher2,refael_moore, Laflorencie2005,lin, Hoyos2007}, conclusively show that concomitant to results obtained for clean systems,
 the disorder averaged EE between a subsystem of size $\ell$ and its complement scales logarithmically with system size, $S(\ell) = \left(c_{\rm eff}/3\right)\ln\ell+\rm{const}$. However, the ``effective" central charge $c_{\rm eff}$ that controls this logarithmic divergence is different from its clean counterpart. Thus, a wide variety of systems that play host to an infinite disorder fixed point 
 like the one-dimensional Random Transverse Field Ising Model (RTFIM) at its critical point, the random antiferromagnetic XXZ chain, the random antiferromagnetic spin-S chain at its random singlet phase and the random q-state Potts chain, et al. display this ``universal" 
 logarithmic behavior of the EE.

The logarithmic scaling of the disorder averaged EE elucidated in the above paragraph is predicated on the existence of an infinite disorder fixed point wherein the scaling relation that connects length and time scales are activated or exponential-like.
However, there are situations where the disorder fluctuations drive the system to a conventional random fixed point that support the conventional power-law scaling between length and time scales. The RG flow equations for such a situation was developed in Refs.~\cite{aharony1,aharony2}. The results of these  flow equations were completely in line with the earlier weak coupling RG results of Refs.~\cite{cardy2,dbtrk} and the later holographic results of \cite{Hartnoll:2014cua}. 

In this paper, we calculate the behavior of the holographic EE at such a disordered conventional random critical point. This done by laying recourse to the dual geometrical description \cite{Hartnoll:2014cua,Hartnoll:2007ih,Hartnoll:2008hs,Adams:2011rj,Adams:2012yi,Arean:2013mta,Lucas:2014zea,Ammon:2018wzb} of the disordered field theory. We apply the prescription of Ryu-Takayanagi formula \cite{Ryu:2006bv,Ryu:2006ef,Casini:2011kv,Park:2015dia} to calculate the EE of a disordered Quantum Field Theory (QFT) from the minimal surface area defined in the dual disordered geometry \cite{Hartnoll:2014cua,Hartnoll:2007ih,Hartnoll:2008hs,Adams:2011rj,Adams:2012yi,Arean:2013mta,Lucas:2014zea,Ammon:2018wzb}. For the disordered system the asymptotic geometry is given by an AdS space, whereas the geometry corresponding to the IR regime approaches a Lifshitz geometry. Due to the specific scaling symmetry of the Lifshitz geometry, its dual QFT is believed to become a Lifshitz Field Theory (LFT) which has also the same scale symmetry (see e.g. Refs.~\cite{Kachru:2008yh,Taylor:2008tg,Balasubramanian:2009rx,Ross:2009ar,Korovin:2013bua,Park:2013goa,Park:2013dqa,Park:2014raa,Mozaffara:2016iwm,MohammadiMozaffar:2017nri,MohammadiMozaffar:2017chk}).

In particular, this manuscript looks at the flow of the EE as we traverse from the ultra-violet (UV) CFT to the infra-red (IR) LFT. Its dual gravity and geometric solution were found in Ref.~\cite{Hartnoll:2014cua} in which a disordered source was represented as a massive bulk scalar field. On this known dual geometry, we will study the effect of the randomly disordered source on the UV EE and investigate how the random disordered source modifies the UV EE to the IR one described by the LFT. We will also look into the mutual information between two subsystems when the theory evolves from CFT to LFT.

The rest of the paper is organized as follows:  In Section \ref{SecCFT}, we take account of a two-dimensional CFT deformed by a disordered source. Intriguingly, its dual geometry allows a Lifshitz IR fixed point where the Lifshitz scale symmetry appears. In Section \ref{secEE}, on this background, after investigating the long-range quantum correlation described by the holographic EE, we look into its critical behavior represented by a critical exponent. 
In Section \ref{SecRG}, we study the renormalization group (RG) flow of the holographic EE. In Section \ref{SecMI}, we further investigate how the disordered source affects the mutual information in both UV and IR regions.  The concluding remarks are in section \ref{SecCon} where we also place our results in the context of earlier results in the field of holographic EE and also in the context of disordered quantum many-body systems. Finally, we reiterate several interesting and important properties of a scale invariant theory through the holographic EE formula in the Appendix \ref{AppA}. 
%

\section{The randomly disordered system from Holography}
\label{SecCFT} 

Recently, significant progress was made in understanding the disordered system holographically. It has been shown that when a two-dimensional CFT is deformed by a disorder operator, its contribution is finite even in the IR as well as UV regimes \cite{Hartnoll:2014cua}. The dual geometry of the randomly disordered system has been analytically constructed by using the Poincar\'{e}-Lindstedt resummation technique. Although it is a perturbative geometric solution, it opens a new window to study an IR critical behavior holographically. Interestingly, the dual geometry modified by the disordered fluctuation remains as an AdS space in the UV limit, while in the IR limit it continuously changes into a Lifshitz geometry with a nontrivial dynamical critical exponent. This geometry implies that the dual QFT runs from a CFT to a LFT along the RG flow. Since the dual geometry connects the UV and IR fixed points smoothly, it provides a good playground to study the RG flow caused by a disordered operator. In this section, we briefly review the dual geometry of a randomly disordered system \cite{Hartnoll:2014cua}.

The starting point to describe a disordered fluctuation holographically is a $(2+1)$-dimensional Einstein-scalar gravity described by the following action
\be
S = \fr{1}{16 \pi G} \int d^3 x \sqrt{-g} \ls {\cal R} + \fr{2}{\RL^2} - 2 \pa_a \ph \pa^a \ph - \fr{2m}{\RL^2} \ph^2 \rs ,
\ee
where $\RL$ is an AdS radius and the scalar field corresponds to the source of the disorder.
From now on, we will set $\RL=1$ for simplicity. When $m=-3/4$, the scalar field near the asymptotic AdS boundary can be expanded into
\be
\ph = u^{1/2} \ph_1 (x) + u^{3/2} \ph_2 (x) + \cdots,
\ee
and its dual scalar operator saturates the Harris criterion. In this case, the randomly disordered source can be represented as (see \cite{Hartnoll:2014cua} for the details)
\be
\ph_1 (x) = v \sum_{n=1}^{N-1} A_n \cos\ls k_n x + \g_n \rs ,
\ee
where $k_n=n \D k$ with $\D k =k_0 /N$ and $\g_n$ denotes a random phase uniformly distributed in $(0,2\pi)$. Here $k_0$ indicates the highest mode of the disorder and the disorder amplitude $A_n$ is given by
\be
A_n = 2 \sqrt{S(k_n) \ \D k}\, ,
\ee
where the function $S(k_n)$ controls the correlations of the noise. The random average denoted by $\bra \cdots \ket_R$ is defined in the $N \to \infty$ limit and accounts for the  local Gaussian noise
\be
\bra \ph_1 (x) \ket_R = 0 ~~ {\rm and} ~~ \bra \ph_1 (x) \ph_1 (y)\ket_R = v^2 \d(x-y) .
\ee

Regarding the gravitational back-reaction of the scalar field, the resulting dual geometry of the randomly disordered system becomes up to $v^2$ order \cite{Hartnoll:2014cua}
\be
ds^2 = \fr{1}{u^2} \Big( - \fr{A(u,x)}{F(u)^{\b(v)}}dt^2 + B(u) dx^2 + du^2 \Big) ,
\ee
with
\bea
A(u, x) &=& 1  + v^2 \Big\{ e^{- 2 k_0 u} \big[ (1+2 k_0 u) \ls \,{\ln}\,(2k_0 u)+\g \rs \big] \Big\}  \nn 
&& -  v^2 \big[ 2 k_0 u e^{- 2 k_0 u}   + \,{\ln}\, \big(\fr{2 k_0 u}{\sqrt{1  + k_0^2 u^2} } \big) \big] , \nn
B(u) &=& 1 + v^2 \ls e^{- 2 k_0 u} +\g -1 \rs , \nn
F(u) &=& 1+k_0^2 u^2 , \nn
\b (v) &=& \fr{v^2}{2} ,
\eea
where $\g$ is the Euler constant, $\g \approx 0.577$. Even when $v$ is order one, the gravitational back-reaction of the disordered source is still regular in the entire range of $u$. Interestingly, this geometry smoothly changes from an AdS geometry in the UV {regime} ($u \to 0$)  into a Lifshitz one in the IR {regime} ($u \to \inf$) with the dynamical critical exponent
\be\la{Lifshitz}
z = 1+ \fr{v^2}{2}  .
\ee

Note that the above averaged metric contains only the second order correction $v^2$ caused by disorder fluctuations. Another thing we must notice is that the averaged metric is not singular in the entire range of the radial coordinate $u$. According to the AdS/CFT correspondence, this radial coordinate corresponds to the energy scale of the dual field theory. In the UV region ($u \to 0$), the averaged metric represents a two-dimensional conformal field theory deformed by relevant disorder fluctuations. Along the renormalization group flow, the regularity of the dual geometry implies that the deformed dual CFT smoothly flows to a Lifshitz field theory in the IR limit.  From this fact, we can see that the minimal surface, although it is governed by a non-linear differential equation, also becomes regular only except the UV divergence. The regularity of the minimal surface enables us to take a perturbative expansion in terms of a small disorder strength in the entire range of $u$. Although the averaged metric is not an exact one including all higher order corrections, it is still valid up to second order perturbation if the strength of disorder fluctuations remains small. The same thing is also true for the minimal surface. This fact implies that the small disorder parameter $v$ allows us to find a perturbative solution in the entire range of $u$. Since we have taken into account only the $v^2$ contribution to the averaged metric, the resulting area of the minimal surface is also valid only up to $v^2$ order. In the case of a small disorder strength, the $v^2$ order contribution is the leading correction caused by disorder fluctuations. We can further consider higher order corrections like $v^4$. Though such higher order corrections are also regular and can modify the critical exponent and the minimal surface area \cite{Hartnoll:2014cua}, they are negligible when comparing it with the $v^2$ order contribution for $v \ll 1$. Hereafter, we concentrate on the leading correction of disorder fluctuations occurring at order $v^2$.

\section{Holographic Entanglement Entropy for the Disordered System}
\label{secEE}

In order to study the disorder effect, let us investigate the holographic EE. Since the perturbative expansion with respect to $v$ is applicable in the entire {regime} of the dual geometry, here we focus on the leading correction of $v^2$ order. By using Ryu-Takayangi formula \cite{Ryu:2006bv,Ryu:2006ef}, the EE can be evaluated by calculating the area of the minimal surface extended to the dual geometry, whose boundary have to coincide with the entangling surface dividing a total system into two parts. In order to clarify the entangling surface, we take a subsystem defined in the following interval
\be
- \fr{l}{2} \le x \le \fr{l}{2} .
\ee
The configuration of the minimal surface can be expressed by $u$ as a function of $x$. Using this parametrization, the induced metric on the minimal surface reduces to
\be
ds_{in}^2 = \fr{ 1}{u(x)^2} \big[ B\big(u(x)\big) + u'(x)^2  \big] d x^2  ,
\ee
where the prime indicates a derivative with respect to $x$. Then, the EE is governed by the following action
\be		\la{act:minimalsurf}
S_E  =  \fr{1}{4 G} \int_{- l/2}^{l/2} d x \ \fr{1}{u(x)} \sqrt{B\big(u(x)\big)+ u'(x)^2 }  .
\ee
In order to obtain a unique minimal surface, we should impose two boundary conditions which fix the integral constants of the second order differential equation. Natural boundary conditions are $u(l/2) = 0$ and $u'(0)=0$. Here the first constraint is required because the entangling surface must be located at the boundary $u=0$, while the second is needed to find a smooth minimal surface at the turning point,  $x=0$.

\subsection{In the UV Regime}
\label{secUV}
Before studying the disorder effect in the IR regime, we first investigate the EE in the UV regime satisfying $k_0 l \ll 1$. Above the action is invariant under $x \to - x$, so that we can further reduce the action into
\be 			\la{act:zofx}
S_E = \fr{1}{2 G}    \int_{0}^{l /2-x_*} d x  \ \fr{1}{u(x)} \sqrt{B\big(u(x)\big)+ u'(x)^2 }  .
\ee
where $x_*$ is introduced to denote a UV cutoff in the $x$-direction.
Due to the existence of the small expansion parameter $v$ in the entire {regime}, $u$ can be expanded to be
\be
u(x) = u_0 (x) + v^2 u_2(x) + {\cal O} \ls v^4 \rs .
\ee
Then, $u_0 (x)$ and $u_2 (x)$ satisfy the following equations of motion
\bea			\la{eq:minimalsurface}
0 &=& u_0  u_0''+ u_0'^2+1 , \\
0&=& e^{2 k_0 u_0} \left(\gamma +2 u_0' u_2'+u_2 u_0''+u_0 u_2''-1\right)\nonumber\\
&&-2 k_0 u_0{}^2 u_0''-k_0 u_0+1 .\la{eq:minimalsurface2}
\eea
The boundary conditions discussed above reduces to $u_i = 0$ at $x= \pm l /2$ and $u_i'=0$ at $x=0$.

The solution of the first equation, $u_0$, describes the geodesic curve in a pure AdS$_3$ space \cite{Ryu:2006bv}
\be
u_0 (x) = \half \sqrt{l^2 - 4 x^2} ,
\ee
which automatically satisfies two required boundary conditions. The second equation governs the deformation of the minimal surface caused by the disorder. It does not allow an analytic solution, so that a numerical study is needed to figure out its effect in the IR regime. However, in the UV regime the existence of another small expansion parameter still enables us to look into the EE analytically but perturbatively. In the UV regime where the highest momentum of the disorder is much smaller than the energy scale observing the dual QFT, $k_0 l$ is small and $u_2$ can be further decomposed into
\be
u_2 (x) = u_{20} (x) + \fr{k_0 l}{2}  u_{21} (x)  + {\cal O} \ls k_0^2 l^2\rs .
\ee
The first solution satisfying the equation of motion is given by
\be
u_{20} (x) =  \fr{c_1  l  +  2  c_2  x - \g x^2  }{\sqrt{l^2 - 4 x^2}} .
\ee
The smoothness at $x=0$ fixes $c_2=0$ and the other boundary condition, $u(l/2) = 0$, gives rise to $c_1 = \g l /4$. Using this result, $u_{21} (x)$ satisfying the equation of motion reads
\begin{align}
u_{21} (x) &= \frac{2x^2-l^2}{2l} +\frac{  c_3 l  +  2  c_4   x   }{ \sqrt{l^2-4 x^2}}  
- \frac{ l  x \tan ^{-1}\Big(\frac{2 x}{\sqrt{l^2-4 x^2}}\Big) }{2  \sqrt{l^2-4 x^2}}. 
\end{align}
Imposing again two required boundary conditions, we finally obtain $c_4 = 0$ and $c_3 = \pi l /8$.

Now, let us introduce a UV cutoff, $\e$, in the $u$-direction which is associated with the UV cutoff, $x_*$, introduced previously. Using the perturbative solutions we found, they are related by
\be
x_*  =  \lb 1 - \ls \fr{\pi}{8} k_0  l  +\g \rs v^2 \rb \fr{\e^2}{l} + {\cal O} \ls \e^3 \rs .
\ee
Substituting these results into Eq.~\eq{act:minimalsurf} and integrating it perturbatively,
the holographic EE in the UV regime can be written as
\be		
S_E  = - \frac{c}{3}  \,{\ln}\,  \epsilon  + S_L (l,v,k_0),
\ee
with
\begin{align}   \label{SLofUV}
S_L (l,v,k_0) &=  \frac{c}{3}  \,{\ln}\, l
+\frac{c}{6} \ls   \g v^2   \rs  - \frac{c \pi }{24}  v^2  k_0 l , 
\end{align}
We use the relation, 
$c = 3\RL/2G$, where $G$ is the Newton constant in three dimensional gravity and $\RL=1$ is the radius of AdS$_3$. Here, the first part including the UV divergence can be regarded as the short distance correlation across the entangling surface (two boundary points), while $S_L (l,v,k_0)$ corresponds to a long-range quantum correlation between the subsystem and its complement.

In the UV regime, our result in \eqref{SLofUV} shows that the disorder corrections affect the long-range quantum correlation
 and that  the resulting EE starts to decrease linearly as the subsystem size increases.

\subsection{In the IR Regime}
\label{secIR}

In order to understand the disorder effect in the IR regime, we need to know the exact minimal surface configuration extended into the deep interior of the dual geometry. In the interior corresponding the IR regime of the dual field theory, the expansion parameter used in the previous section is not small, so that we can not apply the previous expansion any more. In this section, therefore, we will investigate the disorder effect in the IR regime numerically. Noting that the minimal surface action in Eq.~\eq{act:minimalsurf} does not depend explicitly on $x$, we can find a conserved quantity
\be
H = - \fr{ B\big(u(x)\big)}{u(x) \sqrt{B\big(u(x)\big) +u'(x)^2}} .
\ee
At the turning point satisfying $u'(x)=0$ at $x=0$, it reduces to
\be
H = - \fr{  \sqrt{B_0 }}{u_0} ,
\ee
where $u_0$ and $B_0$ indicate the turning point and the value of $B$ on it. Comparing these two results, we can determine the subsystem size and EE in terms of $u_0$
\bea
l (u_0, v, k_0) &=& 2 \int_0^{u_0} du \fr{u  \sqrt{B_0} }{\sqrt{B} \sqrt{B u_0^2 - B_0 u^2}} , \\ 
S_E (u_0, v, k_0) &=& \fr{c}{3} \int_\e^{u_0} du \fr{u_0 \sqrt{B}  }{u \sqrt{B u_0^2 - B_0 u^2}} ,
\eea
where a UV cutoff in the second integral is introduced for the regularization. Note that unlike the previous section where $l$ was used as a free parameter, above $l$ is determined by three free parameters, $u_0$, $v$ and $k_0$. Since $c$ appears as an overall constant in the EE formula, the exact value of $c$ is not important to understand the qualitative behavior of the EE.

\begin{figure}
\begin{center}
\quad\subfigure[]{\label{fig1a} \includegraphics[angle=0,width=0.38\textwidth]{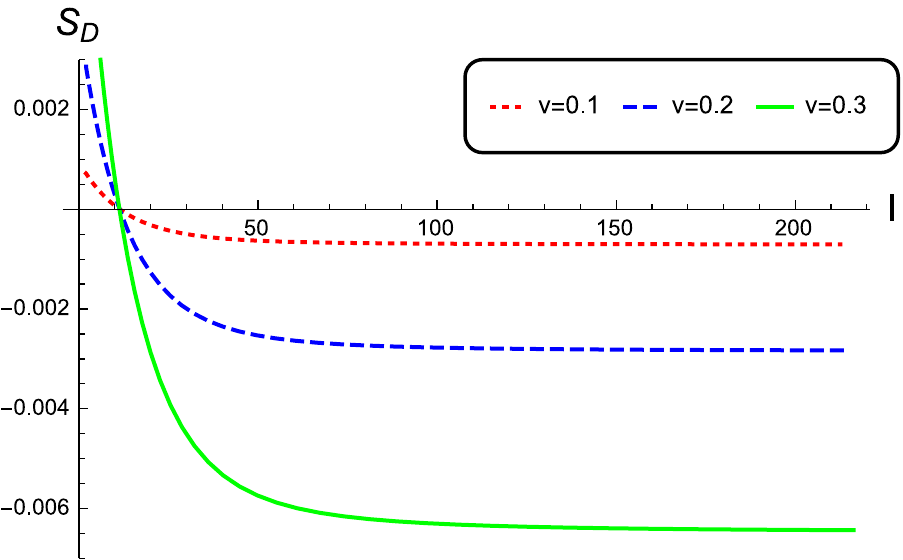}}
\hspace{0cm}
\subfigure[]{\label{fig1b} \includegraphics[angle=0,width=0.38\textwidth]{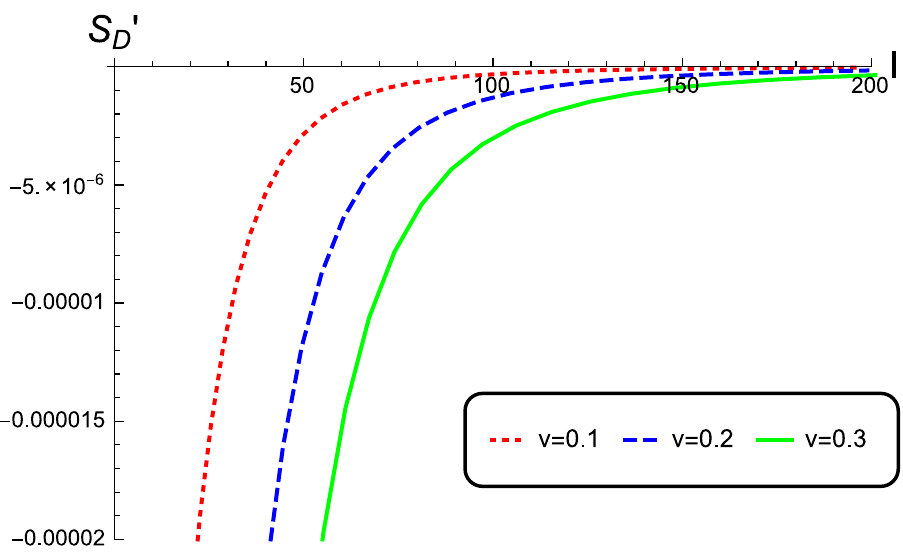}}
\vspace{-0cm}
\caption{\small  The numerical results for $S_D$ and $S_D'$ in terms of $l$, for $v=0.1$ (dotted), $0.2$ (dashed), and $0.3$ (solid). These two results show that $S_D$ in the large $l$ limit approaches to a certain constant value (See also Fig.\,\ref{fig3b}).}
\label{fig1}
\end{center}
\end{figure}

Now, let us concentrate on the long-range quantum correlation in order to look into IR physics
\be \label{SLSD}
S_L = \fr{c}{3} \,{\ln}\, l + S_D  .
\ee
Here, the first term represents the common dependence on the subsystem size, while the second term  corresponds to the disorder effect, which can be obtained from the numerical data of $S_E$, that
\be\label{SESD}
S_E = \fr{c}{3} \,{\ln}\, \fr{l}{\e}+ S_D  \, .
\ee
From now on, we take a specific UV cutoff, $\e = 10^{-10}$,  for the numerical calculation.
If changing the UV cutoff, the value of $S_{E}$ must be changed. However, the qualitative behavior  described by the RG flow is not affected as mentioned before, so that we use $S_D$ and $S_L$ to describe the physical properties.
Fig.\,\ref{fig1a}  shows that the value of $S_D$ in the large $l$ limit converges into a certain constant. This fact becomes clear in Fig.\,\ref{fig1b} where the slope of $S_D$ also approaches to zero. These results strongly indicate that a new scale invariant theory arises at the IR critical point ($l \to \infty$). 

Due to the existence of the scale invariance at the IR critical point, it would be interesting to check whether the IR EE can show a universal critical behavior represented as a critical exponent. Assuming that $S_D$ is given by a polynomial of $l$, $v$ and $k_0$, then near the IR critical point it can be expanded into
\be		\la{exp:IRform}
S_D = S_{\infty} + \frac{C}{ l^{ \d} } + \cdots  .
\ee
Above the ellipsis means higher order corrections, ${\cal O} \ls l^{-\s} \rs$ with $\s>\d$, and can be ignored in the $l \to \infty$ limit. 

\begin{figure}
\begin{center}
\vspace{-0.0cm}
\hspace{-0.0cm}
{\label{fig2a}\includegraphics[angle=0,width=0.375\textwidth]{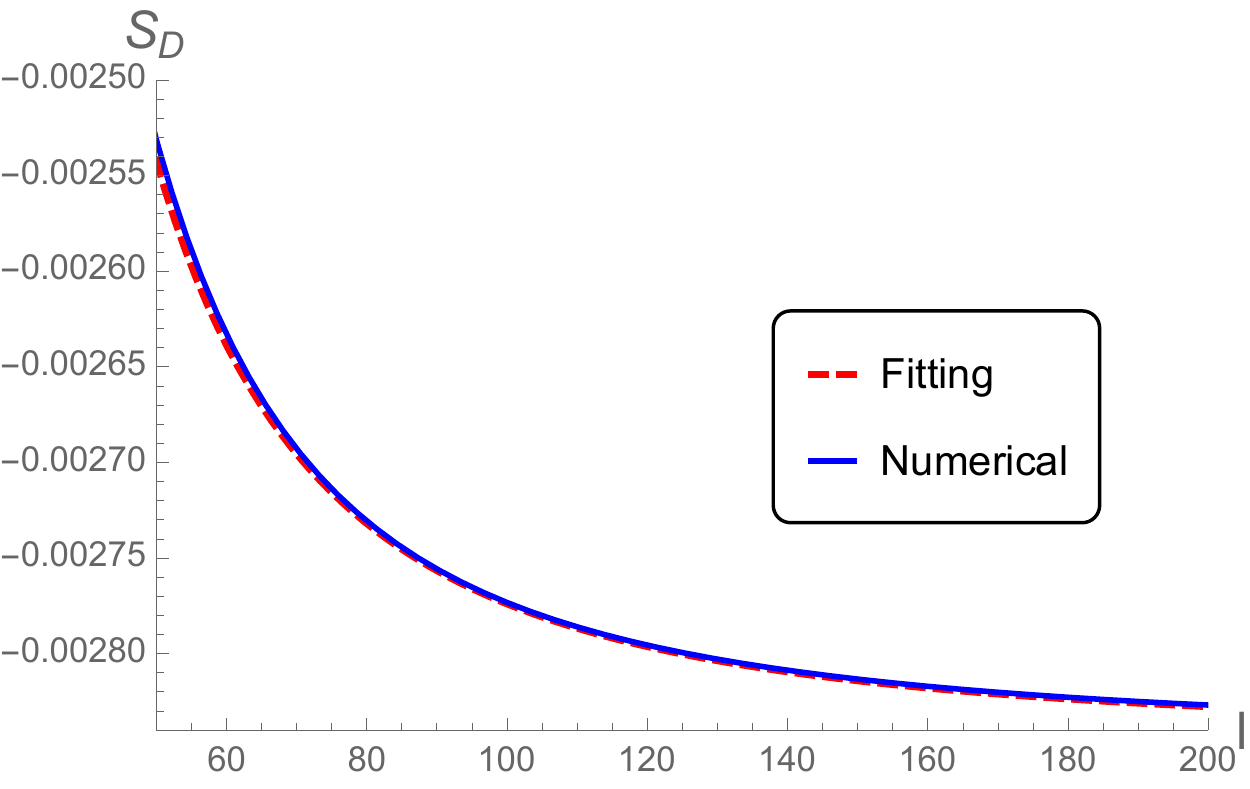}}
\vspace{-0cm}
\caption{\small  
Numerical result of $S_D$ in terms of $l$ (solid curve) and fitted by Eq.~\eq{exp:IRform} (dashed line) with $S_\infty = -0.00284$, $C=1.146$ and $\d=2.107$ for $k_0=0.1$, $v=0.2$ and $c=1$.
 \label{fig2}}
\end{center} 
\end{figure}

In order to check the above expected form, we fit numerical data of $S_D$ with Eq.~\eq{exp:IRform}.
Fig.\,\ref{fig2} shows in the large $l$ limit that numerical data of $S_D$ can be well fitted by Eq.~\eq{exp:IRform} with $S_\infty = -0.00284$, $C=1.146$ and $\d=2.107$ for $k_0=0.1$, $v=0.2$ and $c=1$. In addition, the negative value of $S_D$ implies that the disorder effect reduces the long-range quantum correlation.

\begin{figure}
\begin{center}
\vspace{-0cm}
\subfigure[~$S_{\infty}(v)$ is independent of $k_0$]{\label{fig3a} \includegraphics[angle=0,width=0.35\textwidth]{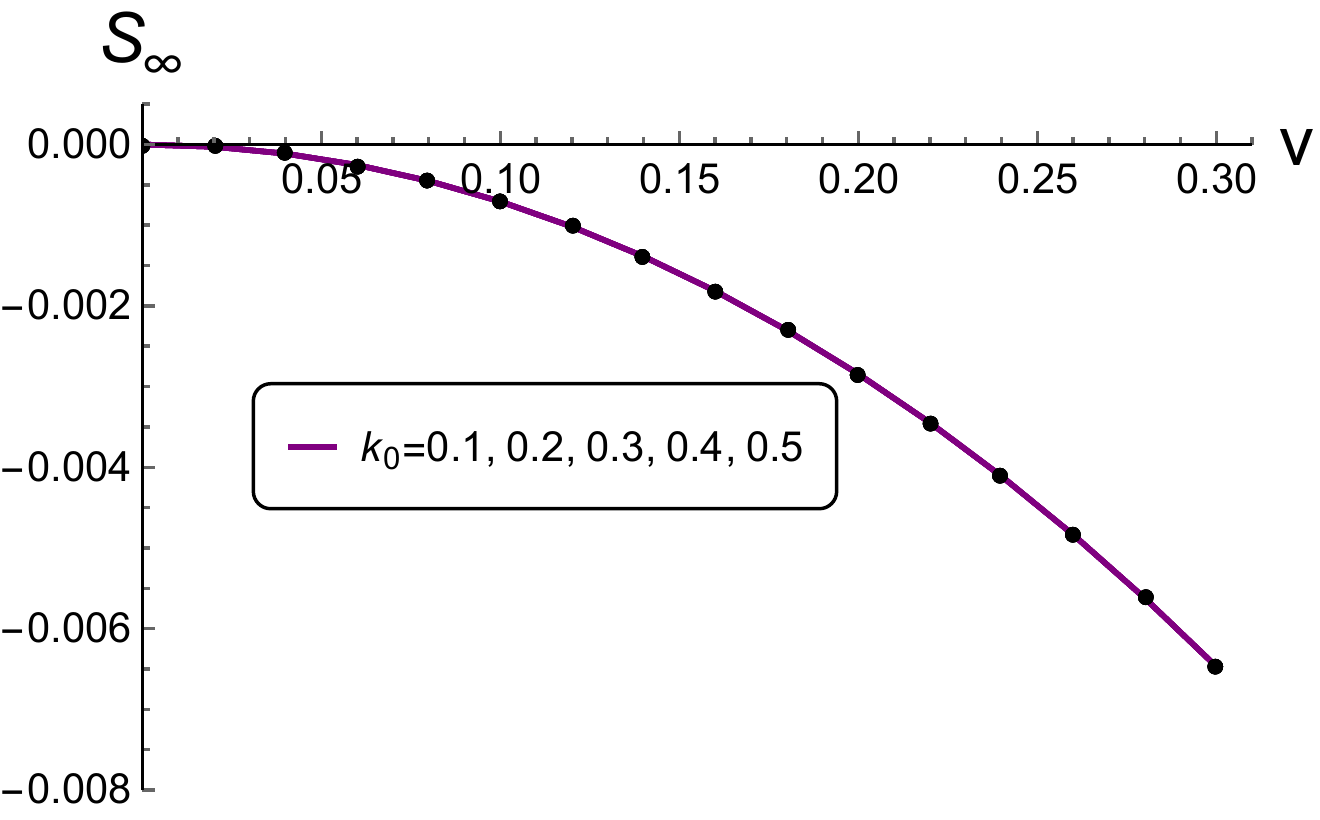}}\\
\subfigure[~$S_L(v)$ is independent of $k_0$]{\label{fig3b} \includegraphics[angle=0,width=0.35\textwidth]{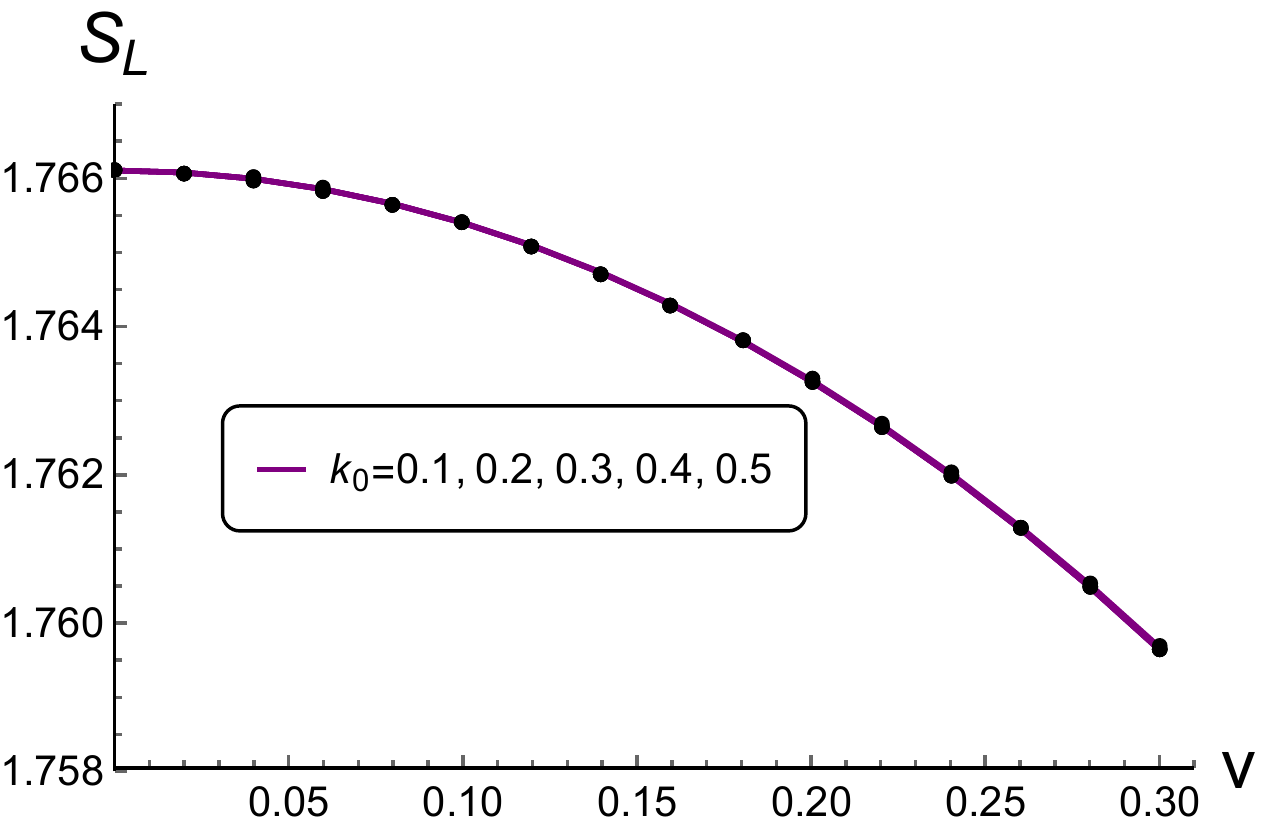}}  \\
\quad\subfigure[~$C(v)$ for different values of $k_0$]{\label{fig3c} \includegraphics[angle=0,width=0.35\textwidth]{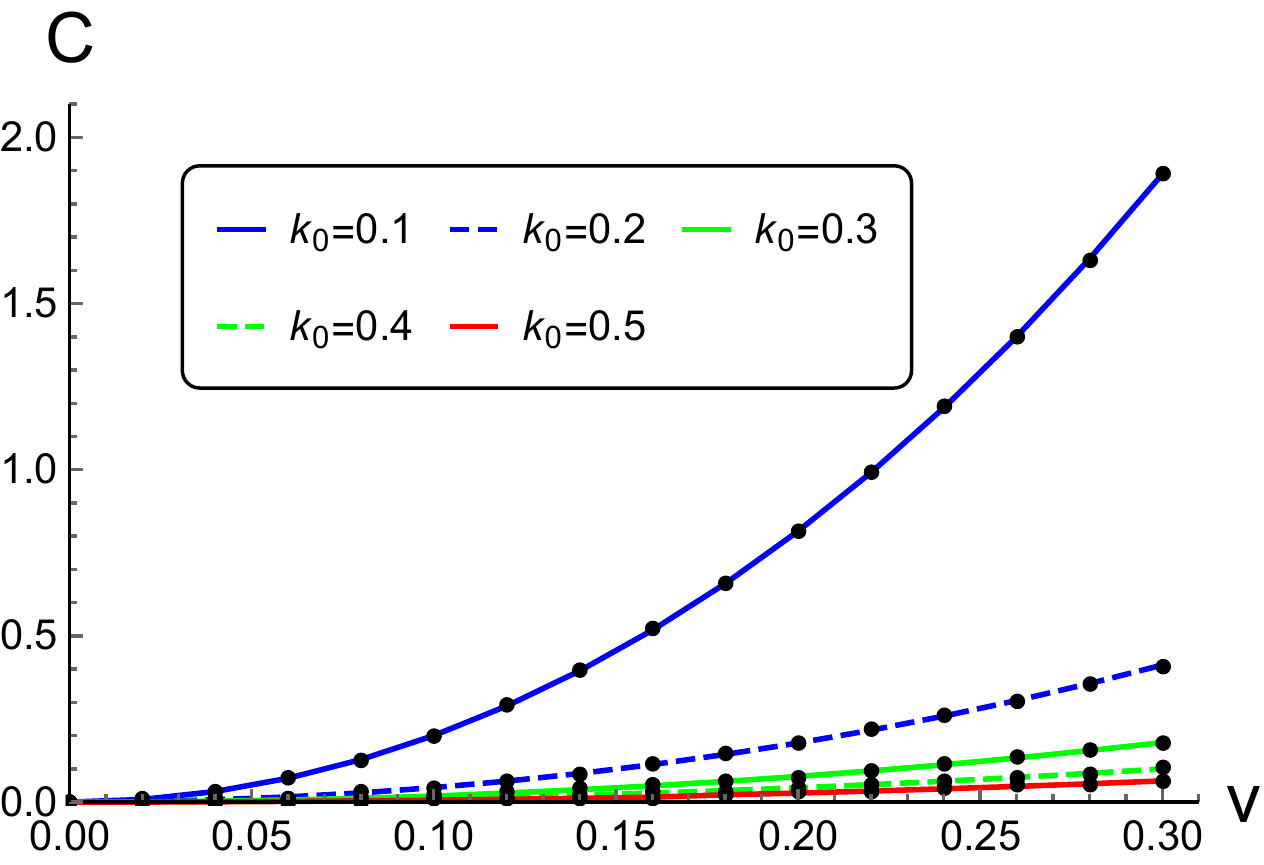}}\\
\quad\subfigure[~$\d(v)$ for different values of $k_0$]{\label{fig3d} \includegraphics[angle=0,width=0.35\textwidth]{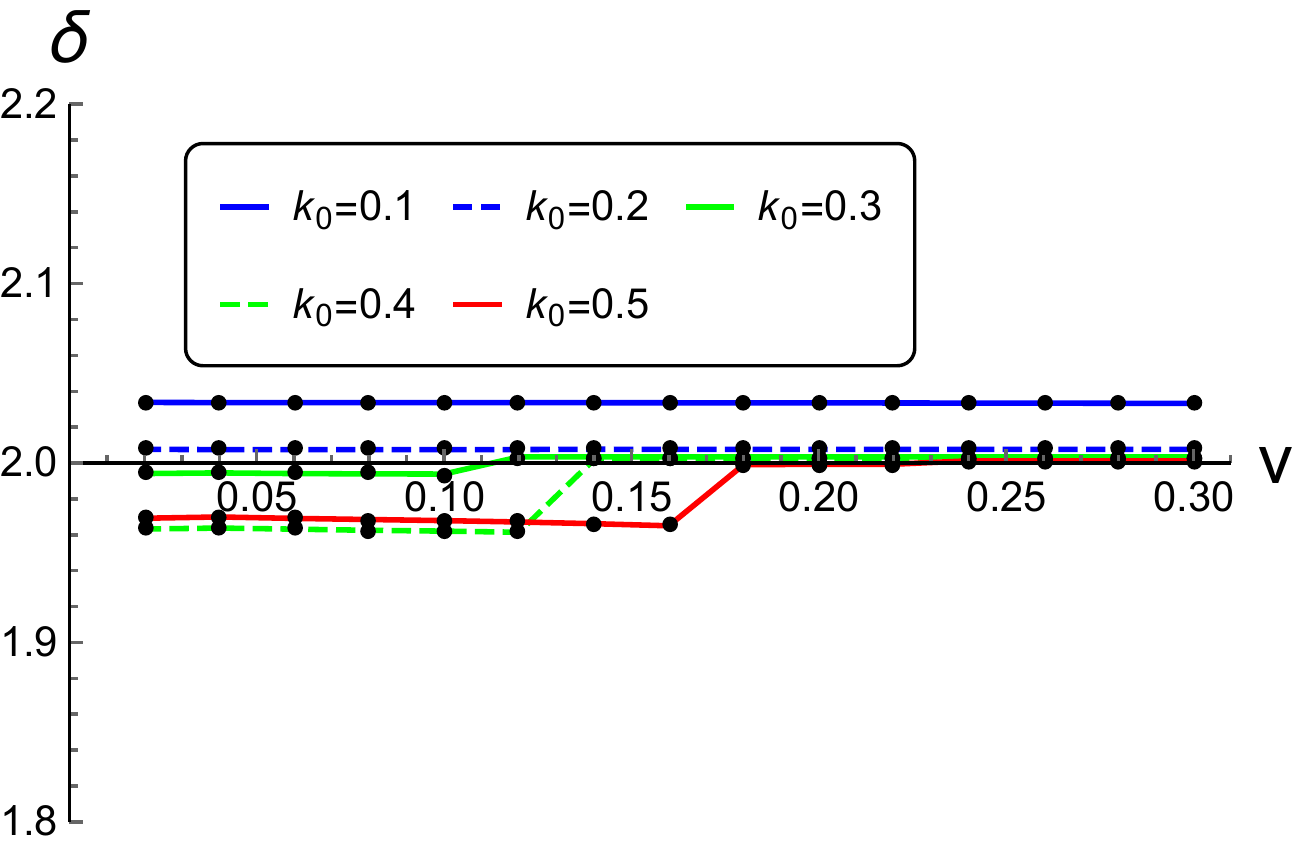}}
\caption{\small  Coefficients of the fitting function in Eq.~\eq{exp:IRform} with $l=200$, $c=1$ and  $k_0$ varies from $0.1$ to $0.5$. $S_{\infty}(v) < 0$ in (a) but $S_L(v) >0$ in (b), which implies that the disorder weakens the long-range correlation. (b) $S_L(v)$ and (c) $C(v)$  show continuous functions, whereas (d) gives an almost constant $\d(v)$ around $2$, which may be interpreted as a critical exponent of the EE near the IR fixed point. The small jumps of $\d$ around $v=0.15$ seems to be a numerical error for $k_0>0.3$.  }
\label{fig3}
\end{center}
\end{figure}

In order to study how the disorder deformation affects the long-range correlation behavior and what the critical behavior of the EE is near the IR fixed point, we extract information about $S_\infty$, $S_L$, $C$ and $\d$ depending on $v$ in Fig.\,\ref{fig3}, where we use $l=200$, $c=1$ and $k_0$ varies from $0.1$ to $0.5$.
The values of $S_{\infty}$ in Fig.\,\ref{fig3a} and $C$ in Fig.\,\ref{fig3c} are given by well-behaved continuous functions. When the amplitude of the disorder $v$ increases $S_\infty$ decreases, while $C$ increases. In general, $\d$ in Eq.~\eq{exp:IRform} is given by a function of the intrinsic parameters of the theory, $k_0$ and $v$. However, the resulting $\d$ in Fig.\,\ref{fig3d} shows an almost constant value with a small deviation for $v>0.3$. In this case, the small deviation seems to be caused by a numerical error. If so, the result in Fig.\,\ref{fig3d} indicates that the first correction to the IR EE is suppressed by a critical exponent $\d$ which is independent of $v$.

There are several remarks before closing this section. First, the critical exponent $\d$ seems to be universal as it depends only very weakly on the amplitude of the disordered fluctuation which is one of the intrinsic parameters of the UV microscopic field theory. Second, this result is obtained by taking into account just $v^2$ order disorder deformation, so that if one further considers higher order corrections like $v^4$, the critical exponent we found can be modified. Lastly, although we obtained a critical exponent of the EE in a small $v$ region, it is not clear whether such a well-defined critical exponent for the long-range EE always appears in the IR critical point, for example, even in the large $v$ limit. When we applied our numerical program to the case with a large value of $v$, our program crashed similar to the previous work in \cite{Hartnoll:2014cua}. Therefore, it remains as an interesting and important issue to check whether there still exists a universal critical exponent in the IR EE even for a large $v$. 

\section{RG flow of Holographic Entanglement Entropy}  
\label{SecRG}

In this section, we encapsulate the results of Sec.~\ref{secUV} and Sec.~\ref{secIR} in terms of RG flow of the EE with the subsystem size $\ell$ acting as the flow-parameter. This description in the language of the RG flow allows us to define the notion of an effective central charge $c_{\rm eff}$ which helps to place our results in context of the well known c-theorem  \cite{Casini:2004bw,Myers:2010tj,Casini:2012ei} for the even dimensional theories and the so-called F-theorem which is its generalization to its odd dimensional counterparts \cite{Myers:2010tj,Jafferis:2011zi}.

In section \ref{secUV}, we have shown in \eqref{SLofUV} that near the UV fixed point, the disorder corrections to EE starts to decrease linearly as the subsystem size increases. In section \ref{secIR}, we have shown in \eqref{exp:IRform} that the disorder effect near the IR fixed point behaves as $ \sim l^{-\d}$ in which the exponent does not crucially depend on the strength of the disordered source. In Fig.\,\ref{fig2}.  we list the numerical data of $S_D$ depending on $l$ in the solid curve and plot the fitted function by Eq.~\eq{exp:IRform} with dashed line. In the following, we will study how the EE of the dual QFT changes along the RG flow \cite{Swingle:2009bg,Taylor:2016aoi,Kim:2016jwu,Kim:2017lyx} .

As explained in Appendix \ref{AppA}, the scale invariance is indispensable in the study of critical behavior. Now, let us deform a UV CFT with a specific relevant operator. Denoting its source (or coupling constant) as $\m$ and its conformal dimension as $\D_s$, the modified EE near the UV fixed point has the following form
\be
S_E =  \fr{c}{3} \,{\ln}\,\fr{l}{\e}   + S_D \ls \m l^{\D_s} \rs ,
\ee
where $S_D$ denotes the contribution caused by the deformation. The nontrivial $l$ dependence of the last term usually breaks the scale symmetry and causes a nontrivial RG flow. In the holographic EE context, when $\e$ is fixed, $1/l$ can be regarded as an RG scale. However, due to the ambiguity associated with the regularization scheme, it is not clear what value one should assign to the UV cutoff and how it is related to the traditional  regularization scheme of a QFT. Although $S_E$ relies on the regularization scheme, its derivative does not. 

This is because the renormalization scheme must be independent of the choice of an UV cutoff.
 This fact becomes manifest when we take into account the RG flow of the EE 
\be    \la{res:RGflowhee}
 \fr{d S_E}{d \,{\ln}\, l} = \fr{c}{3} + \fr{d S_D (\m l^{\D_s})}{d \,{\ln}\, l} .
\ee 
In this case, the scaling behavior of the dimensionful coupling $\m$ is determined by the subsystem size $l$. Since the value of $\m$ generally depends on the energy scale detecting the theory, we can identify the inverse of the subsystem size as the energy scale. Then, the above RG equation describes how the EE is modified under the change of the coupling.

 There exists another interesting and physical interpretation for this RG flow. Let us decompose the EE into
\be			\la{res:LRCinUV}
S_E \equiv  - \frac{ c {\cal A}}{6}  \,{\ln}\,  \epsilon  + S_L  \ls \m l^{\D_s} \rs,
\ee
with
\be
 S_L  \ls \m l^{\D_s} \rs = \frac{ c {\cal A}}{6}  \,{\ln}\,  l  + S_D  \ls \m l^{\D_s} \rs.
\ee
Here the first term including $\,{\ln}\,  \epsilon$ represents a short distance correlation across the entangling surface which gives rise to the area law, ${\cal A}=2$. On the other hand, $S_L$ is independent of the UV cutoff $\e$, which corresponds to the shortest length scale, and can be interpreted as the long-range quantum correlation between the subsystem and its complement. For a two-dimensional LFT, the RG flow of the EE can be well described by the long-range correlation defined above due to the following relation
\be\label{ceff}
\fr{d S_E}{d \,{\ln}\, l} =  \fr{d S_L}{d \,{\ln}\, l} =\frac{c_{\text {eff}}}{3} .
\ee
In this case, the EE $S_L$ caused by the long-range correlation is regularization scheme independent up to a constant term which does not affect the RG flow. When regarding a higher dimensional theory, however, we must be careful \cite{Park:2015dia}. Since the power-law divergence of a higher dimensional theory prohibits us from decomposing the EE naively into short and long-range quantum correlations, an appropriate renormalization procedure is required before extracting a long-range quantum correlation \cite{Taylor:2016aoi}.

When the RG flow of the deformed QFT meets a critical point at a certain IR energy scale, what happens? Generally the scaling symmetry is restored at such a critical point, so the QFT again becomes a scale invariant theory. Due to this fact,  we can expect that the EE is again described by Eq.~\eq{res:heescalinginv} at the IR fixed point. In the following, we will show that the disordered source causes a nontrivial RG flow from a CFT at UV to a LFT at an IR fixed point and that the corresponding holographic EE has the same form at those two fixed points. We will also show that difference in the EE at the two fixed points is encoded into a sub-leading term that is controlled by an exponent that is 
universal at least in the limit of weak disorder. 

\begin{figure}
\begin{center}
\vspace{-0.0cm}
\hspace{-0.0cm}
{\label{fig4a}\includegraphics[angle=0,width=0.38\textwidth]{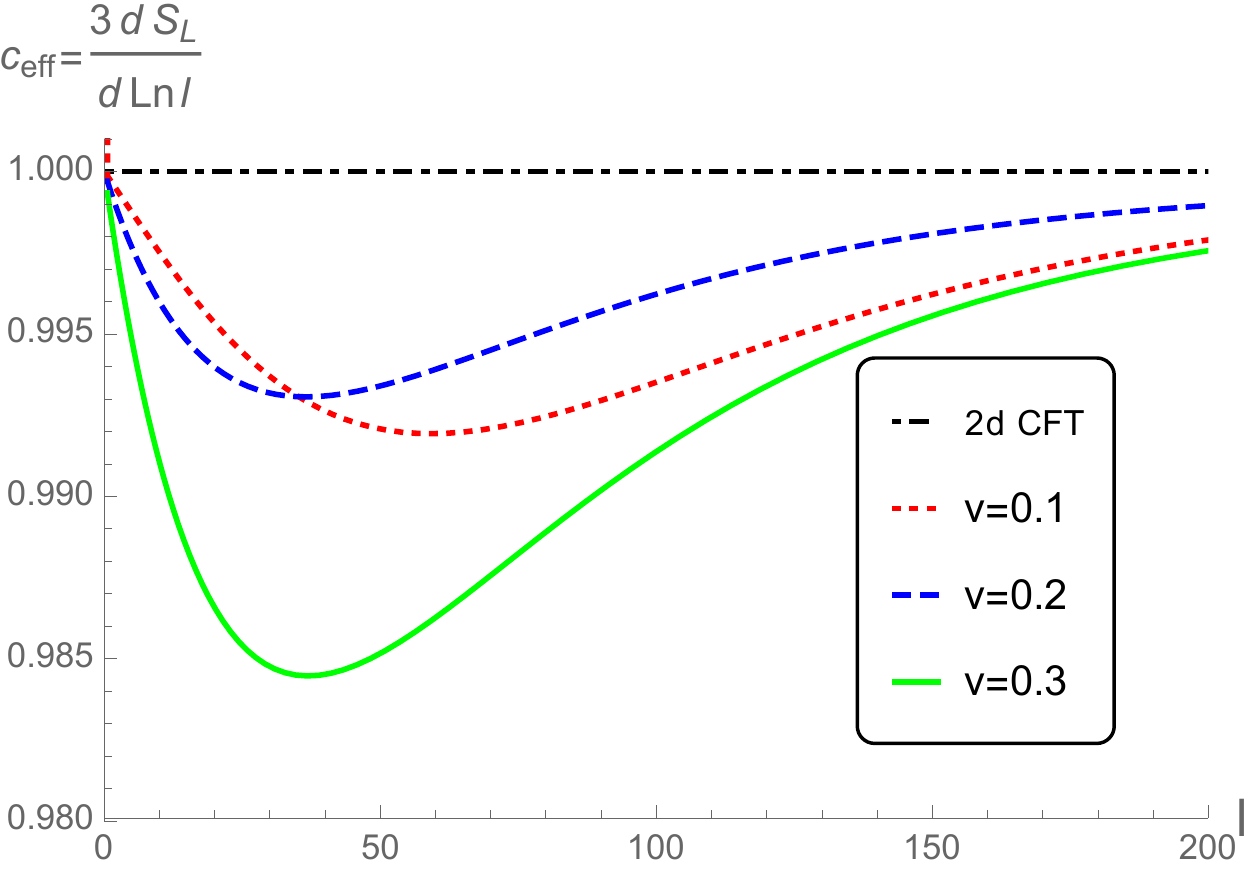}} 
\vspace{0.0cm}
\vspace{-0cm}
\caption{\small 
The change of the long-range quantum correlation $S_L$ in terms of the disorder system relying on the subsystem size $l$, for $k_0=0.1$ and $c=1$, as well as for $v=0.1$ (dotted), $0.2$ (dashed), and $0.3$ (solid).
They can be considered as the RG running of the effective central charge $c_{\text {eff}} \equiv   \fr{3 d S_L}{d \,{\ln}\, l} $. 
The dot-dashed line is a guide to the eye and depicts a case where  $c_{\text{eff}}=1$ does not run.
\label{fig4} }
\end{center}
\end{figure}

The change of the long-range quantum correlation along the RG flow is depicted in Fig.\,\ref{fig4}. It shows that the long-range quantum correlation at the IR Lifshitz fixed point is still proportional to $\,{\ln}\, l$ as a two-dimensional scale invariant theory should do.
Notice that it can also be considered as the RG running of the effective change $c_{\text{eff}}$ in \eqref{ceff}.

At UV CFT and IR Lifshitz fixed points, the theory becomes scale invariant. Due to this scale invariance, the EE at those critical points shows a simple logarithmic behavior relying on the subsystem size. In the EE context, the RG flow can be parametrized by the subsystem size. Using this parametrization, the EE near the UV fixed point decreases linearly along the RG flow. Near the IR fixed point, it still decreases but slowly along the RG flow. In this case, the rate of decrease approaches to zero as the subsystem size goes to the infinity or in the IR limit. In this case, the long-range correlation can be parametrized by the subsystem size by means of an exponent. Intriguingly, via numerical analysis we find that this exponent controlling the vanishing of the  long-range correlation near the IR fixed point is universal for weak disorder. In other words, the critical Lifshitz exponent $z=1+v^2/2$ is very weakly dependent on the disorder strength. It would be interesting to study the IR behavior of the long-range correlation for other models and to find the similar universal critical exponent. 

\section{Holographic Mutual information in the Randomly Disordered System}
\label{SecMI}

In order to understand the IR physics of the randomly disordered system, let us further investigate the mutual information. The mutual information is defined as \cite{Casini:2004bw,Casini:2006ws,Cardy:2013nua,Fischler:2012uv,Swingle:2010jz,Wolf:2008}
\be
I(A,B) = S_E (A) + S_E (B) - S_E (A \cup B) .
\ee
By definition the mutual information has no UV cutoff dependence, so it can measure the long-range quantum correlation. For more details, let us define an operator ${\cal O}_A$ which is in a subsystem A. Then, a two point correlation function between two operators living in different subsystems, $A$ and $B$, can be denoted by
\be
C({\cal O}_A, {\cal O}_B) = \bra {\cal O}_A {\cal O}_B \ket - \bra {\cal O}_A \ket  \bra {\cal O}_B \ket .
\ee
Interestingly, it was known that these two concepts are related to each other \cite{Wolf:2008,Headrick:2010zt,MolinaVilaplana:2011xt}
\be
I(A,B) \ge \fr{  \ls \bra {\cal O}_A {\cal O}_B \ket - \bra {\cal O}_A \ket  \bra {\cal O}_B \ket \rs^2}{2 |{\cal O}_A |^2 |{\cal O}_B|^2 }  \ge 0.
\ee
This relation shows that the mutual information is larger than the square of the two point correlation function. Thus, the mutual information plays a role as the upper bound of a two point correlation function. In addition, the last inequality implies a non-negativity of the mutual information. From this relation, we can see that the vanishing mutual information indicates the absence of the quantum correlation between two disjoint subsystems. Due to this reason,
together with the EE,
the mutual information was used as an indicator of the phase transition in the holographic setup
(see eg. \cite{Nishioka:2006gr,Klebanov:2007ws,Nishioka:2009un,Cai:2012sk,Cai:2017ihd}).

\begin{figure}
\begin{center}
\vspace{+0cm}
\subfigure[]{\label{fig5a} \includegraphics[angle=0,width=0.375\textwidth]{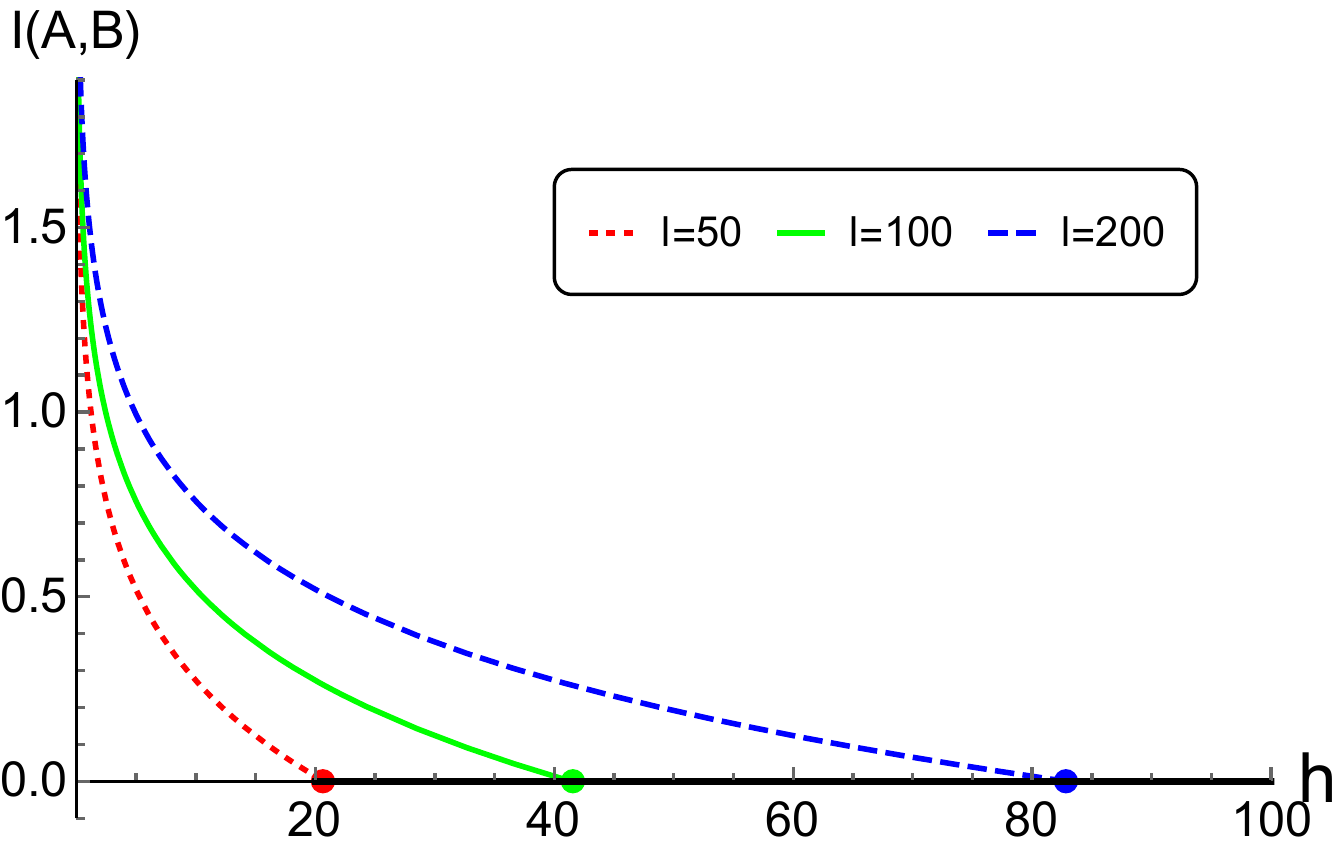}}
\hspace{0.0cm}
\subfigure[]{\label{fig5b}  \includegraphics[angle=0,width=0.375\textwidth]{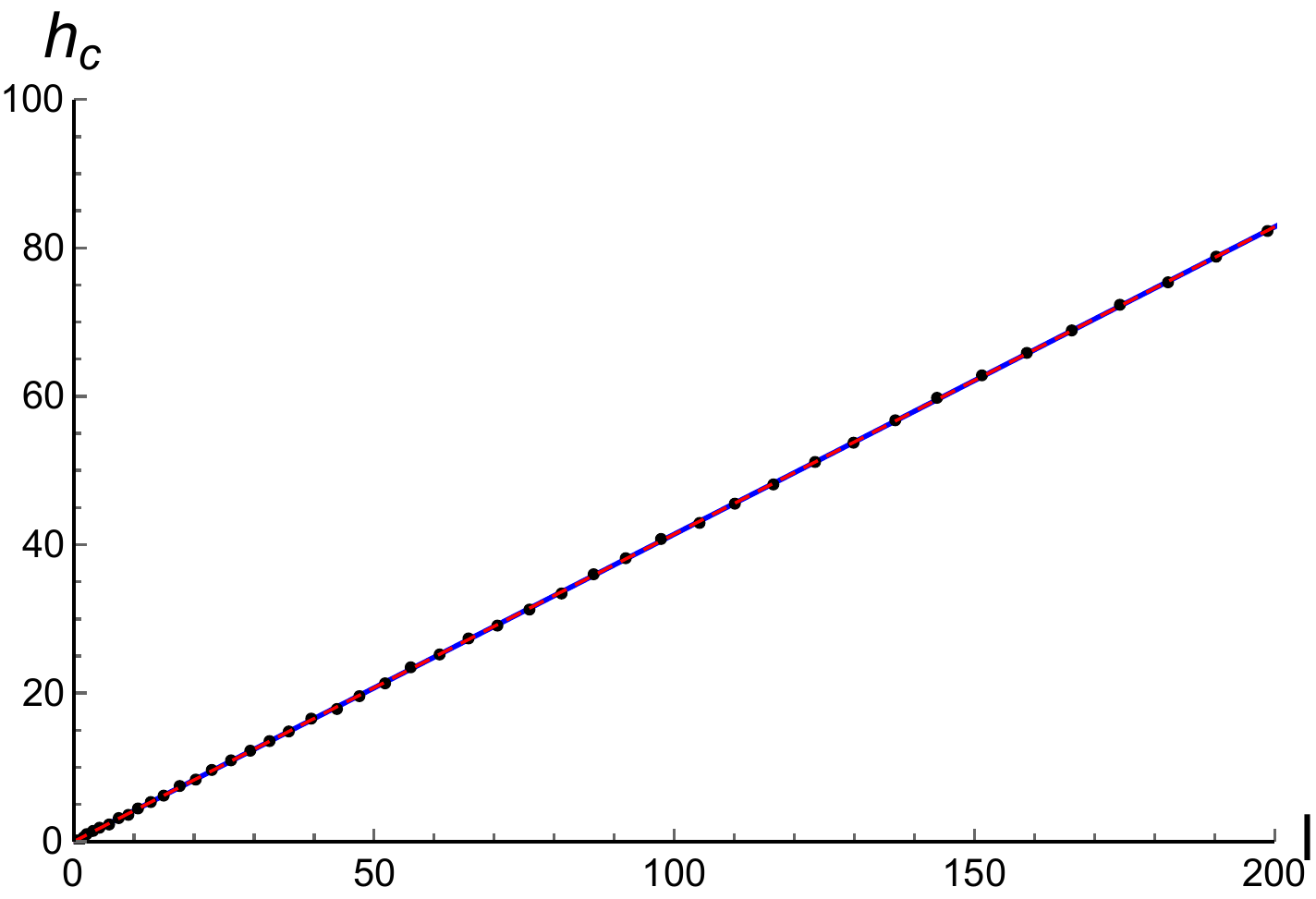}}
\caption{\small (a) The mutual information in terms of the internal distance $h$ and 
(b) the critical distance $h_c$ in terms of $l$ between two subsystems where we have taken $k=0.1$ and $v=0.1$. In (b), the critical distance has a small deviation from the CFT result (the detail is depicted in Fig. \ref{fig6}).}
\label{fig5}
\end{center}
\end{figure}

Now we set two subsystems $A$ and $B$ to have the same size $l$ and $h$ indicates the distance between two closest boundaries of the two subsystems. From the dual geometry of the randomly disordered system and \eqref{SLofUV}, we can easily derive the mutual information which in the UV limit ($l \sim h \ll 1$)
 is written as the following perturbative form
\be
I(A,B)  = \fr{c}{3} \,{\ln}\, \fr{l^2}{h (2 l +h)}  + \frac{c \pi }{12} v^2 k_0  h  +\cdots ,
\ee
Then, the critical distance $h_c$ when the phase transition occurs at $I(A,B)=0$ is given by
\be			\la{hUV}
h_c = (\sqrt{2} -1) l\Big(1 + \frac{\sqrt{2}  \pi }{16}v^2 k_0 l  \Big)+ {\cal O} \ls l^3\rs .
\ee
Below the critical distance ($h <h_c$), the mutual information becomes positive. This implies that the two subsystems have nontrivial quantum correlations. In other words, the union of the subsystems has a lower EE than the EE sum of the subsystems. At the critical distance ($h = h_c$), the mutual information vanishes [see Fig.~\ref{fig5a}] and the quantum correlation between the subsystems also disappear. Above the critical distance ($h > h_c$), since there is no quantum correlation between the subsystem, the EE of the union of the subsystem, $S_E (A \cup B)$, must be the same as the EE sum of each subsystem, $S_E (A) + S_E ( B)$. For the CFT without the disorder ($v=0$), the critical distance is linearly proportional to the subsystem size. On the other hand, the existence of the disorder increases the mutual information and the critical distance slightly in the UV regime. For a CFT, the critical distance is linearly proportional to the subsystem size in the entire RG scale. In the UV regime of the randomly disordered system, the similar behavior also happens with small higher order corrections [see Eq.~\eq{hUV}]. This is because the UV regime of the randomly disordered system can be regarded as a small deformation of the UV CFT.

In Fig.~\ref{fig5a}, we depict the mutual information relying on $h$ at a given subsystem size and  plot how the critical distance depends on the subsystem size in Fig.~\ref{fig5b}. Interestingly, the numerical result in Fig.~\ref{fig5b} shows a linearly increasing behavior, which can be well fitted by
\be
h_c \sim   0.41 \,  l  - 0.004  ,
\ee
where we take $c=1$, $v=0.1$, $k_0 = 0.1$, and $\d=2.034$. Can we understand the critical distance relying on the subsystem size linearly even in the IR regime? In order to discuss the mutual information in the IR limit, let us first take into account the fitting function in Eq.~\eq{exp:IRform} which explains the IR EE of the disordered system well. Using this fitting function,  the mutual information in the IR limit ($l \sim h \gg 1$) can be represented as
\be		\la{res:IRminformation}
I(A,B) =  \fr{c}{3} \,{\ln}\, \fr{l^2}{h (2 l + h)} + C \Big[ \fr{2}{l^{\d}} - \fr{1}{h^{\d}}  - \fr{1}{(2 l + h)^{\d}} \Big] + \cdots ,
\ee
where we set the subsystem sizes to be $l$. In this result, the UV cutoff dependence and the constant part, $S_{\infty}$, of the EE do not appear because they are exactly cancelled. Since these terms are associated with the short distance correlation of the quantum entanglement, the mutual information can be regarded as the measure observing the long-range correlation, as mentioned before.

\begin{figure}
\begin{center}
\vspace{+0cm}
{\label{fig6a} \includegraphics[angle=0,width=0.38\textwidth]{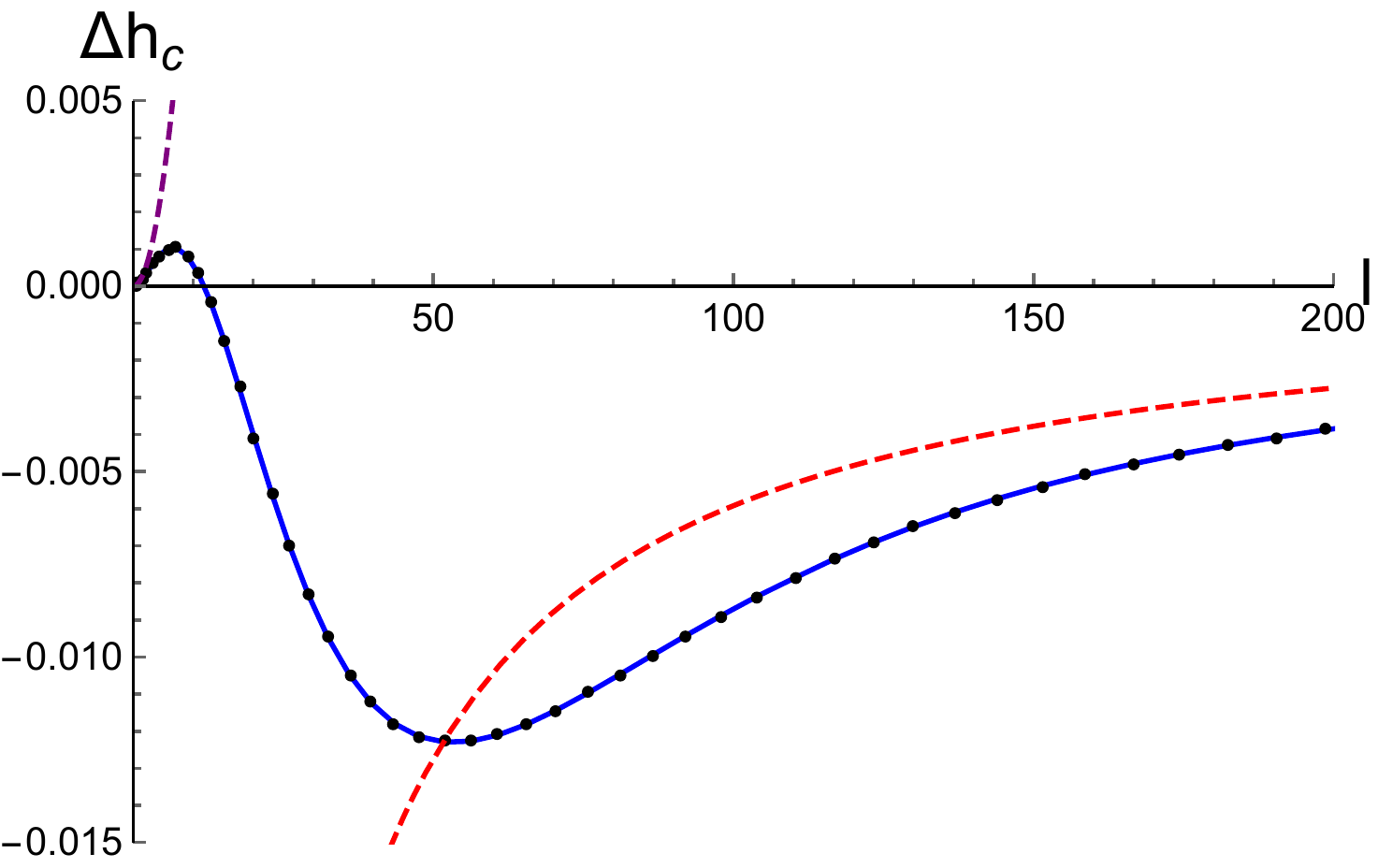}}
\hspace{0.0cm}
\caption{\small The numerical result for relative critical distance $\Delta h_c\equiv h_c-(\sqrt{2}-1)l$  in terms of the internal distance $l$ (Blue curve). 
The purple dashed line is from the analytical result of UV limit \eqref{hUV},
and the red dashed line is from the analytical result of  IR limit \eqref{hIR}.
}
\label{fig6}
\end{center}
\end{figure}

The mutual information in Eq.~\eq{res:IRminformation} has a critical distance of $h_c$ where the mutual information vanishes. Above this critical distance, the long-range quantum correlation between two disjoint systems disappears. In the IR region, the second term in Eq.~\eq{res:IRminformation} is relatively small. Thus, the critical distance satisfying $I(A,B) =0$ can be expressed as
\be
h_c = a l + b l^\l + \cdots ,
\ee
where $\l <1$. The leading term linearly proportional to $l$ is required to make the logarithmic term of the mutual information in Eq.~\eq{res:IRminformation} vanish. After substituting this ansatz into Eq.~\eq{res:IRminformation} and expanding it, we can determine the critical distance as
\begin{align}\label{hIR}
h_c  &=  (\sqrt{2}-1) l \nn 
&-  \fr{3 \big[ (\sqrt{2}+1)^\d + (\sqrt{2} -1)^\d-2 \big] }{2\sqrt{2}  c} \frac{C}{l^{\d-1}} + \cdots.
\end{align}
Noting that $\sqrt{2}-1 \approx 0.41$ and $\l= - (\d-1)$. Similar to the UV case in \eqref{hUV}, the critical distance is still linearly proportional to the subsystem size with a small negative correction. In Fig.~\ref{fig6}, we depict the deviation of the critical distance from the CFT case with $\D h_c = 0$. In the limits having a small and large subsystem sizes, the analytic forms of the critical distance in Eq.~\eq{hUV} and Eq.~\eq{hIR} are well matched to the numerical data in Fig.~\ref{fig6}. This result shows that the critical distance is almost linearly proportional to the subsystem size with a small corrections caused by the disordered source.

Now, let us consider the mutual information in the IR region with a large $l$. The short range correlation can be represented by a small distance between two subsystems ($h \ll 1 \ll l$). In this case, the mutual information reduces to
\begin{align}
I (A,B) & =  \fr{c}{3} \,{\ln}\, \fr{l^2}{h (2 l + h)} + \Big[ S_{\infty} {-}\frac{c  }{6}    \g  v^2  \Big] \nn
& \quad + C \Big[ \frac{2}{l^{\d}} - \frac{1}{(2 l + h)^{\d}} \Big] + \frac{ \pi c}{24} v^2 k_0 h + \cdots \nn
&\approx   \fr{c}{3} \,{\ln}\, \fr{l}{2 h}   +  S_{\infty} {-}\frac{c }{6}    \g v^2 \, .
\end{align}
Note that the leading contribution to the mutual information is independent of the parameters describing the disorder. Since $S_\infty$ decreases as $v$ increases in Fig.~\ref{fig3a}, the above result indicates that the short range quantum correlation decreases as the amplitude of the disorder becomes large. When $h$ increases, the mutual information decreases logarithmically.  For the long-range correlation with $l \gtrsim h\gg 1$, the $h$ dependence of the mutual information at a given $l$ reduces to
\be
I(A,B) \sim  I_{CFT} + C \ls \fr{2}{l^{\d}} - \fr{1}{h^{\d}}  - \fr{1}{(2 l + h)^{\d}} \rs,
\ee
where $ I_{CFT} $ indicates the mutual information of the undeformed CFT. For CFT, the long-range mutual information also decreases logarithmically by $- \,{\ln}\, \lb h (2 l + h) \rb$. The second term above shows the effect of the disordered source in the IR limit.

\vspace{10pt}


\section{Conclusion and Discussion}
\label{SecCon}

In this work, we have holographically investigated the EE of a two-dimensional CFT deformed by a disordered source. In the AdS/CFT contexts, the dual geometry of this disordered CFT has been constructed in \cite{Hartnoll:2014cua}. There are several noticeable points in this dual geometry. The dual scalar field of a disordered source has a finite gravitational back-reaction in the entire {regime}, so that the perturbative expansion in terms of the strength of the disorder is possible even in the IR regime.  Since the disorder is relevant, the UV CFT flows to another  theory in the IR regime. Intriguingly, the disorder considered here allows a Lifshitz fixed point in the IR regime. This Lifshitz fixed point is usually classified by the dynamical critical exponent $z=1+v^2/2$, which in this model is determined by the strength $v$ of the disorder. 
 
In this context, we have shown that effect of these long-range correlations at the IR fixed point can be encoded in a power-law sub-leading term in Eq.~\eq{exp:IRform}, where the Holographic EE vanishes in the limit of infinite sub-system size $\ell$. From Fig.\,\ref{fig3}, the vanishing of the power-law term is controlled by an exponent that numerically seems to be universal at least in the limit of weak-disorder.  Furthermore, since the scaling symmetry is restored at the IR fixed point $\ell\gg 1$, we recover the leading logarithmic dependence on the sub-system size for the holographic EE in Eq.~\eq{SESD}. Notice that the constant $S_{D}|_{\ell\to\infty}=S_{\infty}$ in Fig.\,\ref{fig3a}, which is independent of $k_0$, but monotonically decrease with $v$.  These behaviors match well with the IR Lifshitz  fixed point with the dynamical critical exponent in Eq.~\eq{Lifshitz} that $z = 1+ \fr{v^2}{2}$.

We have also taken account of the quantum EE and its RG flow in Sec.~\ref{SecRG} in order to figure out quantum aspects of the IR physics. To do so, we have decomposed the EE into two parts in Eq.~\eq{res:LRCinUV}, short distance and long-range correlations. The short distance correlation occurs due to the entanglement across the entangling surface, so it is usually associated with the UV divergence of the EE. On the other hand, the UV finite term accounts for the long-range quantum correlation between the subsystem and its complement. As a consequence, the quantum IR physics can be well described by the long-range correlation which is the main point of interest in this study.

We now turn our attention and place our results in the context of earlier results in the holographic study of EE. In the context of holographic EE, it has been proven that the a-type anomaly of an even dimensional theory monotonically decreases along the RG flow due to the unitarity and strong subadditivity of the EE \cite{Casini:2004bw,Myers:2010tj,Casini:2012ei}. In addition, it has been further generalized to an odd dimensional theory where there is no a-type anomaly. The F-theorem has been conjectured as an odd dimensional c-theorem in which the free energy of an odd dimensional QFT decreases monotonically along the RG flow \cite{Myers:2010tj,Jafferis:2011zi}. 
In contrast, the effective central charge $c_{\rm eff}$ as defined in Eq.~\eq{ceff} shows a non-monotonic behavior, (see Fig.~\ref{fig4}). It first decreases along the RG flow and then increases with increasing subsystem size to saturate back to the value of the central charge of the disorder free theory.

The results for the behavior of the EE and concomitantly of the central charge $c_{\rm eff}$ in Eq.~\eq{ceff} that we have obtained near the Lifshitz fixed point is also at odds with that obtained in the context of studying EE in the context of many-body systems that are controlled by an infinite disorder fixed point \cite{Moore2009, Laflorencie2005, Hoyos2007}. In these works, as discussed in the 
introduction, the EE conforms to a form $S(\ell) = \left(c_{\rm eff}/3\right)\ln\ell+\rm{const}$ with an effective central charge $c_{\rm eff}$ that  decreases along the RG flow. This behavior of the EE at the infinite disorder fixed point is explained via the SDRG \cite{Hoyos2007} 
wherein the error of the order of lattice spacing in the placement of effective spins during the RG process contributes corrections to the $\rm{const}$ term thus rendering the $c_{\rm eff}$ universal. This is of course directly in contradiction to our results wherein the $c_{\rm eff}$ shows the non-monotonic behavior encapsulated in Fig.~\ref{fig4}. However, at this juncture, we point out that our results for the Holographic EE were done for a conventional IR fixed point with a finite value of disorder \cite{aharony1, aharony2, cardy2, dbtrk}, whereas the results highlighted in Ref.~\cite{Moore2009, Laflorencie2005, Hoyos2007} have been obtained for an infinite disorder fixed point.

We finally turn our attention to the limits of validity of our approach for the calculation of the EE by using the disordered metric derived in \cite{Hartnoll:2014cua}. Our approach is valid as long as we are in the weak disorder limit $(v\ll 1)$. Thus, in philosophy, our approach is very similar to the effective field theory approaches of disordered systems wherein one integrates out the randomness and work with a disorder averaged field theory from the very outset \cite{cardy2, aharony1}.  Once we leave the limit of small disorder, then atypical or rare-events become important and it would not be appropriate to work with the disorder averaged metric of \cite{Hartnoll:2014cua}. In the strong disorder regime, it would be more appropriate to calculate the EE for each realization of the disorder and then perform an average over the disordered ensembles. This is a much more difficult problem whose resolution we hope to report on in future work. 
For quantum disordered many body systems a numerical efficient implementation of the SDRG in higher dimensions, allows for the  evaluation of the EE at the at the infinite disorder critical point in the 2D RTFIM. It is seen that the EE is given by the area law modified by a logarithmic correction that diverges logarithmically at the critical point \cite{kovacs}. It was further shown that the pre-factor of this logarithmic divergence was a universal number that is independent of the disorder. This result also successfully put to rest conflicting earlier results on the EE in the 2D RTFIM \cite{Yu,Lin}. Now, for higher dimensional disordered system controlled by a conventional disordered fixed point it is possible to calculate the EE by generalizing the disordered metric \cite{Hartnoll:2014cua} following the methodology highlighted in this paper. 
We leave this interesting problem as a future endeavour that can be pursued.


\appendix

\section{Holographic Entanglement Entropy of a Scale Invariant Theory}
\label{AppA}

Let us reiterate several important features of a scale invariant QFT and its dual geometry: The most well-known scale invariant theory is a conformal field theory (CFT).
Another scale invariant theory with reduced symmetry is a Lifshitz  field theory (LFT), which is generally classified by the dynamical critical exponent denoted by $z$. In particular, the conformal symmetry is restored at $z$=1. A LFT shows a specific dispersion relation with $\o \sim p^{z}$. The same scale symmetry can be realized in a Lifshitz geometry. For example, the metric of a three-dimensional Lifshitz geometry is given by
\be
ds^2 = -  \fr{dt^2}{u^{2 z}}  + \fr{dx^2}{u^2}  + \fr{du^2 }{u^2} ,
\ee
where $u$ indicates a radial coordinate of the Lifshitz geometry. The above metric is invariant under the scale transformation:
\be		\la{res:scalingbehavior}
u \to \l u   , \qquad t \to \l^{z} t  \quad {\rm and} \qquad x \to \l x ,
\ee
 From the point of view of symmetry, this Lifshitz geometry can be considered as the dual of a two-dimensional LFT. In the AdS/CFT context, the radial coordinate of the AdS space is identified with the energy scale probing the dual CFT. This identification is useful to connect the RG flow of the condensed matter system to the dual geometry \cite{Rosenhaus:2014nha,Kim:2016ayz,Kim:2016hig,Kim:2017lyx}. Especially, the $u \to 0$ limit corresponds to the UV region. The identification of the radial coordinate of the Lifshitz geometry with the RG scale of the dual LFT entails that the Lifshitz geometry appears in the IR regime after deforming an AdS space with a specific relevant operator, as will be seen.

Now, let us discuss the holographic EE of such scale invariant theories. As explained before, the scale invariant theories have the well defined dual geometries. In such dual geometries the EE can be holographically calculated by applying the Ryu-Takayanagi formula \cite{Ryu:2006bv,Ryu:2006ef}. In the holographic setup, the EE corresponds to the area of a minimal surface which is extended to the dual geometry. In the Lifshitz geometry the holographic EE is expressed as
\be			\la{act:minimalsurface}
S_E = \fr{1}{4G} \int_{-l/2}^{l/2} d x \ \fr{1}{u(x)}  \sqrt{1 + u'(x)^2} ,
\ee
where $l$ denotes the size of a subsystem and the prime means a derivative with respect to $x$. The configuration of the minimal surface is determined by $u(x)$. Note that this formula accounts for the ground state EE. Another point we should note is that the above EE does not depend on the dynamical critical exponent. This is because we do not consider the time evolution of the EE. When time is fixed, the minimal surface representing the holographic EE lives in a spatial section of the dual geometry and the resulting EE becomes independent of the time component of the metric. Since the dual geometries of a CFT and a LFT are distinguishable only in the time direction, their entanglement entropies have no distinction without considering the time evolution of the EE. On the other hand, if we take into account the time evolution of the EE, we must exploit the covariant formalism instead of the Ryu-Takayanagi formula \cite{Hubeny:2007xt}. Since the covariant formulation usually includes information about the time component of the metric, it can distinguish a LFT from a CFT. In this work, we focus only on the time independent case where the scaling behavior of the holographic EE is governed by the scale transformation of spatial coordinates, $u \to \l u$ and $x \to \l x$. One can easily see that the above holographic EE is invariant under this spatial scale transformation.

Through an explicit calculation, it can be shown that the minimal surface describes a geodesic curve represented as $u(x) = \sqrt{l^2/4 - x^2}$. Substituting it into the Ryu-Takayanagi formula, the resulting EE reproduces the known two-dimensional CFT result
\be     \la{res:heescalinginv}
S_E \sim \fr{c}{3} \,{\ln}\,\fr{l}{\e} + {\rm const}  ,
\ee
where $c = \fr{3 \RL}{2 G}$  indicates the central charge and  $\e$ indicates a UV cutoff (or the lattice spacing), respectively. 
$G$ is the Newton constant in three dimensional gravity and $\RL$ is the radius of AdS$_3$. This result can also be understood by the previous spatial scale symmetry. Since the dimensionful parameters scaled under the spatial scaling are $\e$ and $l$, only possible scale invariant combination is given by $l/\e$. After the dimension counting, the allowed terms are the result in Eq.~\eq{res:heescalinginv} and terms having the form of $\ls \e/l \rs^\a$ with a positive power ($\a >0$). Since the UV cutoff is taken to be $\e \to 0$, the terms $\sim \ls \e/l \rs^\a$ automatically vanish. Note that above the constant term is not uniquely determined because of the regularization scheme dependence. Anyway, the lesson from the scale symmetry is that the EE of a scale invariant two-dimensional theory has no nontrivial dependence on $l$ except the logarithmic term.

Although information about the dynamical critical exponent is not involved in the ground state EE, it is not the case for the EE of the excited state. The EE change between the ground and excited states can be described by a relative entropy which is independent a regularization scheme and generally leads to the EE bound \cite{Park:2015hcz}. 
It has been shown that the positivity of the relative entropy leads to the EE bound, and the thermodynamics-like law appears when the EE bound is saturated \cite{Park:2015hcz,Park:2015afa}. The thermodynamics-like law of the EE allows us to reconstruct the AdS dual geometry from data of the boundary conformal field theory (CFT) \cite{Nozaki:2013vta,Lashkari:2013koa,Faulkner:2013ica,Swingle:2014uza,Faulkner:2017tkh}.

Especially, when the EE bound is saturated, it gives rise to the thermodynamics-like law, $\D E = T_E \D S_E$. In this case, the increased EE is directly related to the excitation energy, so that the scaling behavior of the time direction becomes important. For a LFT, the entanglement temperature $T_E$ crucially relies on the dynamical critical exponent
\be		\la{res:entangtem}
T_E \sim \fr{1}{l^{ z}} .
\ee
In order to understand $z$-dependence of the entanglement temperature, we need to recall the scaling behavior in Eq.~\eq{res:scalingbehavior}. The EE is invariant under the scale transformation as mentioned before, and the energy scales as $E \to \l^{- z} E$. Thus, the entanglement temperature must scale as $T_E \to \l^{-z} T_E$. Since the above thermodynamics-like law is regularization scheme independent, the scaling behavior of the entanglement temperature should be explained by the remaining parameter, $l$. As a result, the result in Eq.~\eq{res:entangtem} naturally appears because of the scale invariance of a LFT.

\section*{Acknowledgement}


R. Narayanan would like to thank APCTP for their hospitality during his visit. This work was supported by the Korea Ministry of Education, Science and Technology, Gyeongsangbuk-Do and Pohang City. R. Narayanan was partially supported by National Research Foundation (NRF) funded by MSIP of Korea (Grant No. 2015R1C1A1A01052411). C. Park was also supported by Basic Science Research Program through the National Research Foundation of Korea funded by the Ministry of Education (NRF-2016R1D1A1B03932371).
Y. L. Zhang was supported by the Grant-in-Aid for JSPS international research fellow (18F18315). 




\begin{thebibliography}{99}
 \small
 \bibitem{vojta_review} T.~Vojta,
``Rare region effects at classical, quantum and nonequilibrium phase transitions,''
\href{https://doi.org/10.1088/0305-4470/39/22/R01}{J.\ Phys.\ A {\bf 39}, R143-R205 (2006)}.

\bibitem{harris} A.~B.~Harris,
``Upper bounds for the transition temperatures of generalized Ising models,''
\href{https://doi.org/10.1088/0022-3719/7/17/018}{J.\ Phys.\ C {\bf 7}, 1671 (1974)}.

\bibitem{fisher1}D.~S.~Fisher,
``Random antiferromagnetic quantum spin chains,''
\href{https://doi.org/10.1103/PhysRevB.50.3799}{Phys.\ Rev.\ B {\bf 50},3799 (1994)}


\bibitem{fisher2} D.~S.~Fisher,
``Critical behavior of random transverse-field Ising spin chains,''
\href{https://doi.org/10.1103/PhysRevB.51.6411}{Phys.\ Rev.\ B {\bf 51},6411 (1995)}


\bibitem{Thill_Huse:1995} M.~Thill and D.~A.~Huse,
``Equilibrium behaviour of quantum Ising spin glass,''
\href{https://doi.org/10.1016/0378-4371(94)00247-Q}{Physica A, {\bf 214}, 321 (1995)}.


\bibitem{Rieger_Young:1996}
H.~Rieger  and A.~P.~Young,
``Griffiths singularities in the disordered phase of a quantum Ising spin glass,''
\href{https://doi.org/10.1103/PhysRevB.54.3328}{Phys.\ Rev.\ B {\bf 54},3328 (1996)}

\bibitem{Vojta:2003}
T. ~Vojta 
``Disorder-Induced Rounding of Certain Quantum Phase Transitions,''
\href{https://doi.org/10.1103/PhysRevLett.90.107202}{Phys.\ Rev.\ Lett.\ {\bf 90}, 107202 (2003)}.

\bibitem{Hoyos_Vojta:2008}
J.~A.~Hoyos, T.~Vojta,
``Theory of Smeared Quantum Phase Transitions,''
\href{https://doi.org/10.1103/PhysRevLett.100.240601}{Phys.\ Rev.\ Lett.\ {\bf 100}, 240601 (2008)}.

\bibitem{aizenmann} 
  M.~Aizenman and J.~Wehr,
  ``Rounding of First Order Phase Transitions in Systems With Quenched Disorder,''
   \href{http://dx.doi.org/10.1103/PhysRevLett.62.2503}{Phys.\ Rev.\ Lett.\  {\bf 62}, 2503 (1989)}.

\bibitem{berker} K. Hui and A. N. Berker,
``Random-field mechanism in random-bond multicritical systems,''
\href{https://doi.org/10.1103/PhysRevLett.62.2507}{Phys.\ Rev.\ Lett.\ {\bf 62}, 2507 (1985)}.


\bibitem{berker2} A. N. Berker,
``Critical behavior induced by quenched disorder,''
\href{https://doi.org/10.1016/0378-4371(93)90341-Z}{Physica (Amsterdam) {\bf 194A}, 72 (1993)}.

\bibitem{Anderson} 
  E.~Abrahams, P.~W.~Anderson, D.~C.~Licciardello and T.~V.~Ramakrishnan,
   ``Scaling Theory of Localization: Absence of Quantum Diffusion in Two Dimensions,''
  \href{https://doi.org/10.1103/PhysRevLett.42.673}{Phys.\ Rev.\ Lett.\  {\bf 42}, 673 (1979)}.

\bibitem{SachdevQFT}
S.~Sachdev, ``Quantum Phase Transitions,''
\href{http://www.cambridge.org/9780521514682}{2nd ed.}, Cambridge, England, 2011

\bibitem{osterloh}
A.~Osterloh, L.~Amico, G.~Falci, R.~Fazio,
``Scaling of Entanglement close to a quantum phase transition,''
\href{http://dx.doi.org/10.1038/416608a}{Nature (London) {\bf 416}, 608 (2002)}.

\bibitem{osborne} 
  T.~J.~Osborne and M.~A.~Nielsen,
  ``Entanglement in a simple quantum phase transition,''
  \href{http://dx.doi.org/10.1103/PhysRevA.66.032110}{Phys.\ Rev.\ A {\bf 66}, 032110 (2002)}.

\bibitem{chen} Y.~Chen, P.~Zanardi, Z.~D.~Wang and F.~C.~Zhang,
``Sublattice entanglement and quantum phase transitions in antiferromagnetic spin chains''
\href{https://doi.org/10.1088/1367-2630/8/6/097}{New J. Phys. {\bf 8}, 97 (2006)}.


\bibitem{amico}
  L.~Amico, R.~Fazio, A.~Osterloh and V.~Vedral,
  ``Entanglement in many-body systems,''
  \href{http://dx.doi.org/10.1103/RevModPhys.80.517}{Rev.\ Mod.\ Phys.\  {\bf 80}, 517 (2008)}
  [\href{http://arxiv.org/abs/quant-ph/0703044}{quant-ph/0703044 [QUANT-PH]}].

\bibitem{vidal}
  G.~Vidal, J.~I.~Latorre, E.~Rico and A.~Kitaev,
  ``Entanglement in quantum critical phenomena,''
  \href{http://dx.doi.org/10.1103/PhysRevLett.90.227902}{Phys.\ Rev.\ Lett.\  {\bf 90}, 227902 (2003)}
  [\href{http://arxiv.org/abs/quant-ph/0211074}{quant-ph/0211074}].
\bibitem{Moore2009} 
  G.~Refael and J.~E.~Moore,
  ``Criticality and entanglement in random quantum systems,''
  \href{http://dx.doi.org/10.1088/1751-8113/42/50/504010}{J.\ Phys.\  A:  Math.\ Theor.\ {\bf  42}, 504010 (2009)}.
  [\href{https://arxiv.org/abs/0908.1986}{arXiv:0908.1986[cond-mat.dis-nn]}]



\bibitem{ma_dasgupta_hu}
 S.~K.~Ma, C.~Dasgupta, and C.~K.~Hu,
 ``Random Antiferromagnetic Chain,''
\href{https://doi.org/10.1103/PhysRevLett.43.1434}{Phys.\ Rev.\ Lett.\ {\bf 43}, 1434 (1979)}.

\bibitem{ma} C.~Dasgupta and S.~K.~Ma,
``Low-temperature properties of the random Heisenberg antiferromagnetic chain,''
\href{https://doi.org/10.1103/PhysRevB.22.1305}{Phys.\ Rev.\ B {\bf 22},1305 (1980)}

\bibitem{refael_moore} 
  G.~Refael and J.~E.~Moore,
  ``Entanglement Entropy of Random Quantum Critical Points in One Dimension,''
  \href{http://dx.doi.org/10.1103/PhysRevLett.93.260602}{Phys.\ Rev.\ Lett.\  {\bf 93}, 260602 (2004)}.


\bibitem{Laflorencie2005}  
N.~Laflorencie,  
  ``Scaling of entanglement entropy in the random singlet phase,''
  \href{http://dx.doi.org/10.1103/PhysRevB.76.174425}{Phys.\ Rev.\ B\  {\bf 72}, 140408 (R) (2005)}.
  [\href{https://arxiv.org/abs/cond-mat/0504446}{arXiv: cond-mat/0504446 [cond-mat.str-el]}]


\bibitem{lin}
F. Igl\'{o}i and Y.-C. Lin,
``Finite-size scaling of the entanglement entropy of the quantum Ising chain with homogeneous, periodically modulated and random couplings''
\href{https://doi.org/10.1088/1742-5468/2008/06/P06004}{J.\ Stat.\ Mech.\ P06004 (2008)}.

\bibitem{chiara}
  G.~De Chiara, S.~Montangero, P.~Calabrese and R.~Fazio,
   ``Entanglement entropy dynamics in Heisenberg chains,''
  \href{https://doi.org/10.1088/1742-5468/2006/03/P03001}{J.\ Stat.\ Mech.\  {\bf 0603}, P03001 (2006)}
  [\href{http://arxiv.org/abs/cond-mat/0512586}{cond-mat/0512586}].

\bibitem{Hoyos2007}  
J. A.~Hoyos, A. P.~Vieira, N.~Laflorencie and E.~Miranda,
  ``Correlation amplitude and entanglement entropy in random spin chains,''
  \href{http://dx.doi.org/10.1103/PhysRevB.76.174425}{Phys.\ Rev.\ B\  {\bf 76}, 174425 (2007)}.
  [\href{https://arxiv.org/abs/0704.0951}{arXiv:0704.0951[cond-mat.dis-nn]}]



\bibitem{aharony1}
  V.~Narovlansky and O.~Aharony,
  ``Renormalization Group in Field Theories with Quantum Quenched Disorder,''
  \href{https://doi.org/10.1103/PhysRevLett.121.071601}{Phys.\ Rev.\ Lett.\  {\bf 121}, no. 7, 071601 (2018)}
  [\href{http://arxiv.org/abs/arXiv:1803.08529}{arXiv:1803.08529 [cond-mat.str-el]}].
  



\bibitem{aharony2} 
  O.~Aharony and V.~Narovlansky,
  ``Renormalization group flow in field theories with quenched disorder,''
  \href{https://doi.org/10.1103/PhysRevD.98.045012}{Phys.\ Rev.\ D {\bf 98}, no. 4, 045012 (2018)}
  [\href{http://arxiv.org/abs/arXiv:1803.08534}{arXiv:1803.08534 [hep-th]}].


\bibitem{dbtrk} T.~R.~Kirkpatrick and D.~Belitz, 
``Long-Range Order versus Random-Singlet Phases in Quantum Antiferromagnetic Systems with Quenched Disorder", 
\href{https://doi.org/10.1103/PhysRevLett.76.2571}{Phys. Rev. Lett. {\bf 76}, 2571}
Erratum [\href{https://doi.org/10.1103/PhysRevLett.78.1197}{Phys. Rev. Lett. {\bf 78}, 1197 (1997)}].

 
\bibitem{cardy2} 
D.~Boyanovsky and J.~L.~Cardy, 
``Critical behavior of m-component magnets with correlated impurities'', 
\href{https://doi.org/10.1103/PhysRevB.26.154}{Phys. Rev. B {\bf 26}, 154, (1982)}.
 

\bibitem{Hartnoll:2014cua}
  S.~A.~Hartnoll and J.~E.~Santos,
  ``Disordered Horizons: Holography of Randomly Disordered Fixed Points,''
  \href{http://dx.doi.org/10.1103/PhysRevLett.112.231601}{Phys.\ Rev.\ Lett.\  {\bf 112}, 231601 (2014)}
  [\href{http://arxiv.org/abs/arXiv:1402.0872}{arXiv:1402.0872 [hep-th]}].

\bibitem{Hartnoll:2007ih}
  S.~A.~Hartnoll, P.~K.~Kovtun, M.~Muller and S.~Sachdev,
   ``Theory of the Nernst effect near quantum phase transitions in condensed matter, and in dyonic black holes,''
 \href{http://dx.doi.org/10.1103/PhysRevB.76.144502}{Phys.\ Rev.\ B {\bf 76}, 144502 (2007)}
  [\href{http://arxiv.org/abs/arXiv:0706.3215}{arXiv:0706.3215 [cond-mat.str-el]}].

\bibitem{Hartnoll:2008hs}
  S.~A.~Hartnoll and C.~P.~Herzog,
   ``Impure AdS/CFT correspondence,''
   \href{http://dx.doi.org/10.1103/PhysRevD.77.106009}{Phys.\ Rev.\ D {\bf 77}, 106009 (2008)}
   [\href{http://arxiv.org/abs/arXiv:0801.1693}{arXiv:0801.1693 [hep-th]}].


\bibitem{Adams:2011rj}
  A.~Adams and S.~Yaida,
   ``Disordered holographic systems: Functional renormalization,''
    \href{http://dx.doi.org/10.1103/PhysRevD.92.126008}{Phys.\ Rev.\ D {\bf 92}, no. 12, 126008 (2015)}
  [\href{http://arxiv.org/abs/arXiv:1102.2892}{arXiv:1102.2892 [hep-th]}].

\bibitem{Adams:2012yi}
  A.~Adams and S.~Yaida,
  ``Disordered holographic systems: Marginal relevance of imperfection,''
    \href{http://dx.doi.org/10.1103/PhysRevD.90.046007}{Phys.\ Rev.\ D {\bf 90}, no. 4, 046007 (2014)}
  [\href{http://arxiv.org/abs/arXiv:1201.6366}{arXiv:1201.6366 [hep-th]}].

\bibitem{Arean:2013mta}
  D.~Arean, A.~Farahi, L.~A.~Pando Zayas, I.~S.~Landea and A.~Scardicchio,
  ``Holographic superconductor with disorder,''
    \href{http://dx.doi.org/10.1103/PhysRevD.89.106003}{Phys.\ Rev.\ D {\bf 89}, no. 10, 106003 (2014)}
  [\href{http://arxiv.org/abs/arXiv:1308.1920}{arXiv:1308.1920 [hep-th]}].

\bibitem{Lucas:2014zea}
  A.~Lucas, S.~Sachdev and K.~Schalm,
   ``Scale-invariant hyperscaling-violating holographic theories and the resistivity of strange metals with random-field disorder,''
    \href{http://dx.doi.org/10.1103/PhysRevD.89.066018}{Phys.\ Rev.\ D {\bf 89}, no. 6, 066018 (2014)}
  [\href{http://arxiv.org/abs/arXiv:1401.7993}{arXiv:1401.7993 [hep-th]}].

\bibitem{Ammon:2018wzb}
  M.~Ammon, M.~Baggioli, A.~JimÃ©nez-Alba and S.~Moeckel,
   ``A smeared quantum phase transition in disordered holography,''
\href{http://doi.org/10.1007/JHEP04(2018)068}{JHEP {\bf 1804}, 068 (2018)}
\href{http://arxiv.org/abs/arXiv:1802.08650}{arXiv:1802.08650 [hep-th]}.
  


\bibitem{Ryu:2006bv}
  S.~Ryu and T.~Takayanagi,
   ``Holographic derivation of entanglement entropy from AdS/CFT Correspondence,''
   \href{http://dx.doi.org/10.1103/PhysRevLett.96.181602}{Phys.\ Rev.\ Lett.\  {\bf 96}, 181602 (2006)}
  [\href{http://arxiv.org/abs/hep-th/0603001}{hep-th/0603001}].

\bibitem{Ryu:2006ef}
  S.~Ryu and T.~Takayanagi,
   ``Aspects of Holographic Entanglement Entropy,''
  \href{http://dx.doi.org/10.1088/1126-6708/2006/08/045}{JHEP {\bf 0608}, 045 (2006)}
  [\href{http://arxiv.org/abs/hep-th/0605073}{hep-th/0605073}].

\bibitem{Casini:2011kv}
  H.~Casini, M.~Huerta and R.~C.~Myers,
   ``Towards a derivation of holographic entanglement entropy,''
    \href{http://dx.doi.org/10.1007/JHEP05(2011)036}{JHEP {\bf 1105}, 036 (2011)}
  [\href{http://arxiv.org/abs/arXiv:1102.0440}{arXiv:1102.0440 [hep-th]}].

\bibitem{Park:2015dia}
  C.~Park,
   ``Logarithmic Corrections to the Entanglement Entropy,''
  \href{http://dx.doi.org/10.1103/PhysRevD.92.126013}{Phys.\ Rev.\ D {\bf 92}, no. 12, 126013 (2015)}
  [\href{http://arxiv.org/abs/arXiv:1505.03951}{arXiv:1505.03951 [hep-th]}].


  

  
  

\bibitem{Kachru:2008yh}
  S.~Kachru, X.~Liu and M.~Mulligan,
   ``Gravity duals of Lifshitz-like fixed points,''
  \href{http://dx.doi.org/10.1103/PhysRevD.78.106005}{Phys.\ Rev.\ D {\bf 78}, 106005 (2008)}
  [\href{http://arxiv.org/abs/arXiv:0808.1725}{arXiv:0808.1725 [hep-th]}].

\bibitem{Taylor:2008tg}
  M.~Taylor,
  ``Non-relativistic holography,''
  \href{http://arxiv.org/abs/arXiv:0812.0530}{arXiv:0812.0530 [hep-th]}.

\bibitem{Balasubramanian:2009rx}
  K.~Balasubramanian and J.~McGreevy,
   ``An Analytic Lifshitz black hole,''
    \href{http://dx.doi.org/10.1103/PhysRevD.80.104039}{Phys.\ Rev.\ D {\bf 80}, 104039 (2009)}
  [\href{http://arxiv.org/abs/arXiv:0909.0263}{arXiv:0909.0263 [hep-th]}].

\bibitem{Ross:2009ar}
  S.~F.~Ross and O.~Saremi,
   ``Holographic stress tensor for non-relativistic theories,''
  \href{http://dx.doi.org/10.1088/1126-6708/2009/09/009}{JHEP {\bf 0909}, 009 (2009)}
  [\href{http://arxiv.org/abs/arXiv:0907.1846}{arXiv:0907.1846 [hep-th]}].

\bibitem{Korovin:2013bua}
  Y.~Korovin, K.~Skenderis and M.~Taylor,
  ``Lifshitz as a deformation of Anti-de Sitter,''
   \href{http://dx.doi.org/10.1007/JHEP08(2013)026}{JHEP {\bf 1308}, 026 (2013)}
  [\href{http://arxiv.org/abs/arXiv:1304.7776}{arXiv:1304.7776 [hep-th]}].

\bibitem{Park:2013goa}
  C.~Park,
  ``Notes on the holographic Lifshitz theory,''
  \href{http://dx.doi.org/10.1155/2014/917632}{Adv.\ High Energy Phys.\  {\bf 2014}, 917632 (2014)}
  [\href{http://arxiv.org/abs/arXiv:1305.6690}{arXiv:1305.6690 [hep-th]}].

\bibitem{Park:2013dqa}
  C.~Park,
  ``Massive quasinormal mode in the holographic Lifshitz Theory,''
  \href{http://dx.doi.org/10.1103/PhysRevD.89.066003}{Phys.\ Rev.\ D {\bf 89}, no. 6, 066003 (2014)}
  [\href{http://arxiv.org/abs/arXiv:1312.0826}{arXiv:1312.0826 [hep-th]}].

\bibitem{Park:2014raa}
  C.~Park,
  ``Review of the holographic Lifshitz theory,''
  \href{http://dx.doi.org/10.1142/S0217751X1430049X}{Int.\ J.\ Mod.\ Phys.\ A {\bf 29}, no. 24, 1430049 (2014)}.


\bibitem{Mozaffara:2016iwm}
  M.~Reza Mohammadi Mozaffar, A.~Mollabashi and F.~Omidi,
   ``Non-local Probes in Holographic Theories with Momentum Relaxation,''
  \href{http://dx.doi.org/10.1007/JHEP10(2016)135}{JHEP {\bf 1610}, 135 (2016)}
  [\href{http://arxiv.org/abs/arXiv:1608.08781}{arXiv:1608.08781 [hep-th]}].

\bibitem{MohammadiMozaffar:2017nri}
  M.~R.~Mohammadi Mozaffar and A.~Mollabashi,
   ``Entanglement in Lifshitz-type Quantum Field Theories,''
   \href{http://dx.doi.org/10.1007/JHEP07(2017)120}{JHEP {\bf 1707}, 120 (2017)}
  [\href{http://arxiv.org/abs/arXiv:1705.00483}{arXiv:1705.00483 [hep-th]}].


\bibitem{MohammadiMozaffar:2017chk} 
  M.~R.~Mohammadi Mozaffar and A.~Mollabashi,
  ``Logarithmic Negativity in Lifshitz Harmonic Models,''
  \href{https://doi.org/10.1088/1742-5468/aac135}{J.\ Stat.\ Mech.\  {\bf 1805}, no. 5, 053113 (2018)}
   [\href{http://arxiv.org/abs/arXiv:1712.03731}{arXiv:1712.03731 [hep-th]}].


%

\bibitem{Casini:2004bw}
  H.~Casini and M.~Huerta,
   ``A Finite entanglement entropy and the c-theorem,''
 \href{http://dx.doi.org/10.1016/j.physletb.2004.08.072}{Phys.\ Lett.\ B {\bf 600}, 142 (2004)}
  [\href{http://arxiv.org/abs/hep-th/0405111}{hep-th/0405111}].

\bibitem{Myers:2010tj}
  R.~C.~Myers and A.~Sinha,
   ``Holographic c-theorems in arbitrary dimensions,''
    \href{http://dx.doi.org/10.1007/JHEP01(2011)125}{JHEP {\bf 1101}, 125 (2011)}
  [\href{http://arxiv.org/abs/arXiv:1011.5819}{arXiv:1011.5819 [hep-th]}].

\bibitem{Casini:2012ei}
  H.~Casini and M.~Huerta,
  ``On the RG running of the entanglement entropy of a circle,''
    \href{http://dx.doi.org/10.1103/PhysRevD.85.125016}{Phys.\ Rev.\ D {\bf 85}, 125016 (2012)}
  [\href{http://arxiv.org/abs/arXiv:1202.5650}{arXiv:1202.5650 [hep-th]}].


\bibitem{Jafferis:2011zi}
  D.~L.~Jafferis, I.~R.~Klebanov, S.~S.~Pufu and B.~R.~Safdi,
  ``Towards the F-Theorem: N=2 Field Theories on the Three-Sphere,''
 \href{http://dx.doi.org/10.1007/JHEP06(2011)102}{JHEP {\bf 1106}, 102 (2011)}
  [\href{http://arxiv.org/abs/arXiv:1103.1181}{arXiv:1103.1181 [hep-th]}].



 
\bibitem{Taylor:2016aoi}
  M.~Taylor and W.~Woodhead,
   ``Renormalized entanglement entropy,''
  \href{http://dx.doi.org/10.1007/JHEP08(2016)165}{JHEP {\bf 1608}, 165 (2016)}
  [\href{http://arxiv.org/abs/arXiv:1604.06808}{arXiv:1604.06808 [hep-th]}].


\bibitem{Swingle:2009bg}
  B.~Swingle,
   ``Entanglement Renormalization and Holography,''
   \href{http://dx.doi.org/10.1103/PhysRevD.86.065007}{Phys.\ Rev.\ D {\bf 86}, 065007 (2012)}
  [\href{http://arxiv.org/abs/arXiv:0905.1317}{arXiv:0905.1317 [cond-mat.str-el]}].


\bibitem{Kim:2016jwu}
  K.~S.~Kim and C.~Park,
  ``Renormalization group flow of entanglement entropy to thermal entropy,''
  \href{http://dx.doi.org/10.1103/PhysRevD.95.106007}{Phys.\ Rev.\ D {\bf 95}, no. 10, 106007 (2017)}
  [\href{http://arxiv.org/abs/arXiv:1610.07266}{arXiv:1610.07266 [hep-th]}].

\bibitem{Kim:2017lyx}
  K.~S.~Kim, S.~B.~Chung and C.~Park,
 ``An emergent holographic description for the Kondo effect: The role of an extra dimension in a non-perturbative field theoretical approach,''
  \href{http://arxiv.org/abs/arXiv:1705.06571}{arXiv:1705.06571 [hep-th]}.




\bibitem{Casini:2006ws}
  H.~Casini,
  ``Mutual information challenges entropy bounds,''
    \href{http://dx.doi.org/10.1088/0264-9381/24/5/013}{Class.\ Quant.\ Grav.\  {\bf 24}, 1293 (2007)}
  [\href{http://arxiv.org/abs/gr-qc/0609126}{gr-qc/0609126}].

\bibitem{Cardy:2013nua}
  J.~Cardy,
   ``Some results on the mutual information of disjoint regions in higher dimensions,''
    \href{http://dx.doi.org/10.1088/1751-8113/46/28/285402}{J.\ Phys.\ A {\bf 46}, 285402 (2013)}
  [\href{http://arxiv.org/abs/arXiv:1304.7985}{arXiv:1304.7985 [hep-th]}].

\bibitem{Fischler:2012uv}
  W.~Fischler, A.~Kundu and S.~Kundu,
   ``Holographic Mutual Information at Finite Temperature,''
    \href{http://dx.doi.org/10.1103/PhysRevD.87.126012}{Phys.\ Rev.\ D {\bf 87}, no. 12, 126012 (2013)}
  [\href{http://arxiv.org/abs/arXiv:1212.4764}{arXiv:1212.4764 [hep-th]}].

\bibitem{Swingle:2010jz}
  B.~Swingle,
  ``Mutual information and the structure of entanglement in quantum field theory,''
  \href{http://arxiv.org/abs/arXiv:1010.4038}{arXiv:1010.4038 [quant-ph]}.

\bibitem{Wolf:2008}
M.~M. Wolf, F.~Verstraete, M.~B. Hastings, and J.~I. Cirac,
``Area laws in quantum systems: Mutual information and correlations,''
\href{http://dx.doi.org/10.1103/PhysRevLett.100.070502}{Phys.\ Rev.\  Lett. {\bf 100}, 070502(2018)}
[\href{http://arxiv.org/abs/arXiv:0704.3906}{arXiv:0704.3906 [quant-ph]}].

\bibitem{Headrick:2010zt}
  M.~Headrick,
   ``Entanglement Renyi entropies in holographic theories,''
\href{http://dx.doi.org/10.1103/PhysRevD.82.126010}{Phys.\ Rev.\ D {\bf 82}, 126010 (2010)}
  [\href{http://arxiv.org/abs/arXiv:1006.0047}{arXiv:1006.0047 [hep-th]}].
\bibitem{MolinaVilaplana:2011xt}
  J.~Molina-Vilaplana and P.~Sodano,
   ``Holographic View on Quantum Correlations and Mutual Information between Disjoint Blocks of a Quantum Critical System,''
  \href{http://dx.doi.org/10.1007/JHEP10 (2011)011}{JHEP {\bf 1110}, 011 (2011)}
  [\href{http://arxiv.org/abs/arXiv:1108.1277}{arXiv:1108.1277 [quant-ph]}].

\bibitem{Nishioka:2006gr}
  T.~Nishioka and T.~Takayanagi,
   ``AdS Bubbles, Entropy and Closed String Tachyons,''
  \href{http://dx.doi.org/10.1088/1126-6708/2007/01/090}{JHEP {\bf 0701}, 090 (2007)}
  [\href{http://arxiv.org/abs/hep-th/0611035}{hep-th/0611035}].

\bibitem{Klebanov:2007ws}
  I.~R.~Klebanov, D.~Kutasov and A.~Murugan,
   ``Entanglement as a probe of confinement,''
  \href{http://dx.doi.org/10.1016/j.nuclphysb.2007.12.017}{Nucl.\ Phys.\ B {\bf 796}, 274 (2008)}
  [\href{http://arxiv.org/abs/arXiv:0709.2140}{arXiv:0709.2140 [hep-th]}].

\bibitem{Nishioka:2009un}
  T.~Nishioka, S.~Ryu and T.~Takayanagi,
   ``Holographic Entanglement Entropy: An Overview,''
  \href{http://dx.doi.org/10.1088/1751-8113/42/50/504008}{J.\ Phys.\ A {\bf 42}, 504008 (2009)}
  [\href{http://arxiv.org/abs/arXiv:0905.0932}{arXiv:0905.0932 [hep-th]}].

\bibitem{Cai:2012sk}
  R.~G.~Cai, S.~He, L.~Li and Y.~L.~Zhang,
   ``Holographic Entanglement Entropy in Insulator/Superconductor Transition,''
  \href{http://dx.doi.org/10.1007/JHEP07(2012)088}{JHEP {\bf 1207}, 088 (2012)}
  [\href{http://arxiv.org/abs/arXiv:1203.6620}{arXiv:1203.6620 [hep-th]}].

\bibitem{Cai:2017ihd}
  R.~G.~Cai, X.~X.~Zeng and H.~Q.~Zhang,
   ``Influence of inhomogeneities on holographic mutual information and butterfly effect,''
   \href{http://dx.doi.org/10.1007/JHEP07(2017)082}{JHEP {\bf 1707}, 082 (2017)}
   [\href{http://arxiv.org/abs/arXiv:1704.03989}{arXiv:1704.03989 [hep-th]}].
  
  
  
\bibitem{kovacs}  I.~A.~Kov\'{a}cs, F. ~Igl\'{o}i,
``Universal  logarithmic  terms  in the  entanglement  entropy  of  2d,  3d  and  4d  random transverse-field Ising model," 
\href{https://doi.org/10.1209/0295-5075/97/67009}{EPL {\bf 97}, 67009 (2012)}.
[\href{http://arxiv.org/abs/arXiv:1108.3942}{arXiv:1108.3942 [cond-mat.stat-mech]}]

\bibitem{Lin} Y.\ -C.\ Lin, F.~Igl\'{o}i, H.~Rieger, 
``Entanglement  entropy  at infinite  randomness  fixed  points  in  higher  dimensions," 
\href{https://doi.org/10.1103/PhysRevLett.99.147202}{Phys.\ Rev.\ Lett.\ {\bf 99}, 147202 (2007)}.
 [\href{http://arxiv.org/abs/arXiv:0704.0418}{arXiv:0704.0418 [cond-mat.dis-nn]}]
	
\bibitem{Yu} R.~Yu, H.~Saleur, S.~Haas, 
``Entanglement  entropy  in the two-dimensional random transverse field Ising model,"
\href{https://doi.org/10.1103/PhysRevB.77.140402}{Phys. Rev. B {\bf 77}, 140402(R) (2008)}.
 [\href{http://arxiv.org/abs/arXiv:0709.3840}{arXiv:0709.3840 [cond-mat.dis-nn]}]
   
  
  

\bibitem{Kim:2016ayz}
  K.~S.~Kim and C.~Park,
  ``Emergent geometry from field theory: Wilson's renormalization group revisited,''
    \href{http://dx.doi.org/10.1103/PhysRevD.93.121702}{Phys.\ Rev.\ D {\bf 93}, no. 12, 121702 (2016)}
  [\href{http://arxiv.org/abs/arXiv:1604.04990}{arXiv:1604.04990 [hep-th]}].
  
\bibitem{Kim:2016hig}
  K. S. Kim, M. Park, J. Cho, and C. Park,
   ``Emergent geometric description for a topological phase transition in the Kitaev superconductor model,''
    \href{http://dx.doi.org/10.1103/PhysRevD.96.086015}{Phys.\ Rev.\ D {\bf 96}, no. 8, 086015 (2017)}
  [\href{http://arxiv.org/abs/arXiv:1610.07312}{arXiv:1610.07312 [hep-th]}].
 
\bibitem{Rosenhaus:2014nha}
  V.~Rosenhaus and M.~Smolkin,
   ``Entanglement Entropy Flow and the Ward Identity,''
    \href{http://dx.doi.org/10.1103/PhysRevLett.113.261602}{Phys.\ Rev.\ Lett.\  {\bf 113}, no. 26, 261602 (2014)}
  [\href{http://arxiv.org/abs/arXiv:1406.2716}{arXiv:1406.2716 [hep-th]}].

\bibitem{Hubeny:2007xt}
  V.~E.~Hubeny, M.~Rangamani and T.~Takayanagi, ``A Covariant holographic entanglement entropy proposal,''
    \href{http://dx.doi.org/10.1088/1126-6708/2007/07/062}{JHEP {\bf 0707}, 062 (2007)}
  \href{http://arxiv.org/abs/arXiv:0705.0016}{[arXiv:0705.0016 [hep-th]]}.
   
  

\bibitem{Park:2015hcz}
  C.~Park,
  ``Thermodynamic law from the entanglement entropy bound,''
  \href{http://dx.doi.org/10.1103/PhysRevD.93.086003}{Phys.\ Rev.\ D {\bf 93}, no. 8, 086003 (2016)}
  [\href{http://arxiv.org/abs/arXiv:1511.02288}{arXiv:1511.02288 [hep-th]}].


\bibitem{Park:2015afa}
  C.~Park,
  ``Holographic entanglement entropy in the nonconformal medium,''
   \href{http://dx.doi.org/10.1103/PhysRevD.91.126003}{Phys.\ Rev.\ D {\bf 91}, no. 12, 126003 (2015)}
   [\href{http://arxiv.org/abs/arXiv:1501.02908}{arXiv:1501.02908 [hep-th]}].

\bibitem{Nozaki:2013vta}
  M.~Nozaki, T.~Numasawa, A.~Prudenziati and T.~Takayanagi,
 ``Dynamics of Entanglement Entropy from Einstein Equation,''
   \href{http://dx.doi.org/10.1103/PhysRevD.88.026012}{Phys.\ Rev.\ D {\bf 88}, no. 2, 026012 (2013)}
   [\href{http://arxiv.org/abs/arXiv:1304.7100}{arXiv:1304.7100 [hep-th]}].

\bibitem{Lashkari:2013koa}
  N.~Lashkari, M.~B.~McDermott and M.~Van Raamsdonk,
  ``Gravitational dynamics from entanglement `thermodynamics',''
 \href{http://dx.doi.org/10.1007/JHEP04(2014)195}{JHEP {\bf 1404}, 195 (2014)}
 [\href{http://arxiv.org/abs/arXiv:1308.3716}{arXiv:1308.3716 [hep-th]}].

\bibitem{Faulkner:2013ica}
  T.~Faulkner, M.~Guica, T.~Hartman, R.~C.~Myers and M.~Van Raamsdonk,
 ``Gravitation from Entanglement in Holographic CFTs,''
  \href{http://dx.doi.org/10.1007/JHEP03(2014)051}{JHEP {\bf 1403}, 051 (2014)}
  [\href{http://arxiv.org/abs/arXiv:1312.7856}{arXiv:1312.7856 [hep-th]}].

\bibitem{Swingle:2014uza}
  B.~Swingle and M.~Van Raamsdonk,
   ``Universality of Gravity from Entanglement,''
  \href{http://arxiv.org/abs/arXiv:1405.2933}{arXiv:1405.2933 [hep-th]}.

\bibitem{Faulkner:2017tkh}
  T.~Faulkner, F.~M.~Haehl, E.~Hijano, O.~Parrikar, C.~Rabideau and M.~Van Raamsdonk,
   ``Nonlinear Gravity from Entanglement in Conformal Field Theories,''
    \href{http://dx.doi.org/10.1007/JHEP08(2017)057}{JHEP {\bf 1708}, 057 (2017)}
  [\href{http://arxiv.org/abs/arXiv:1705.03026}{arXiv:1705.03026 [hep-th]}].





\end{thebibliography}
\end{document}